\documentclass[12pt, draftclsnofoot, onecolumn]{IEEEtran}
\usepackage{amsfonts}
\usepackage{dsfont}
\usepackage[normalem]{ulem}
\usepackage{setspace}
\usepackage{color}
\usepackage{amssymb}
\usepackage{cite}
\usepackage[cmex10]{amsmath}
\usepackage{algorithm}
\usepackage{algorithmic} % algorithm
\usepackage{array}
\usepackage{mathrsfs}
\usepackage{graphicx}
\usepackage{latexsym}
\usepackage{amscd}
\usepackage{amsfonts}
\usepackage{subfigure}
\usepackage{amsmath,amscd,amssymb,verbatim}
\usepackage{graphics}
\usepackage{amsthm}
\usepackage[utf8]{inputenc}
\usepackage{authblk}
\usepackage{amsmath,amsthm}
\newtheorem{lemma}{\underline{Lemma}}%[section]

\begin{document}
\title{
Reliability-Oriented Resource Allocation for Wireless~Powered~Short~Packet~Communications with Multiple   WPT Sources
%Reliability-Oriented Power Allocation for Multi-Source WPT Enabled Short Packet Communications
} % power transportation
\author[$\dag$]{Ning~Guo, Xiaopeng Yuan, Yulin Hu$^*$,   and Anke Schmeink \vspace{-.2cm}
\thanks{Part of this paper, i.e., the materials in   Section~\ref{sec:power allocation}, have been presented in   IEEE International Symposium on Wireless Communication Systems (ISWCS), October 2022, Hangzhou, China~\cite{Reliability2022Guo}.
%} %submitted to IEEE International Conference on Communications (ICC),  May 2019,  Shanghai, China~\cite{ICC_2019}.}
%\thanks{This work was supported in part by  the NSFC Grant with No. 62101389, DFG Grant with No. SCHM 2643/16}
%\thanks{
N. Guo, X. Yuan and  Y. Hu are
  with    School of Electronic Information, 
  Wuhan University,   430072 Wuhan,
  China and INDA Institute, RWTH Aachen University, 
  52074 Aachen, 
  Germany,    
(email: $ning.guo|yulin.hu$@whu.edu.cn, $yuan$@inda.rwth-aachen.de). $^*$Y. Hu is the corresponding author.
%} \thanks{
A. Schmeink is with   INDA Institute, RWTH Aachen University, 52074 Aachen, Germany.   (email: $schmeink$@inda.rwth-aachen.de).
}
}
  %\emph{Member, IEEE}
   %\vspace{-.5cm}
\maketitle
 \vspace{-1.4cm}
%%%%%%%%%%%%%%%%%%%%%%%%%%%%%%%%%%%%%%%%%%%%%%%%%%%%%%%%%%%%%
\begin{abstract}
 \vspace{-.26cm}
%In this paper, 
We study a multi-source wireless power transfer (WPT)  enabled network supporting  multi-sensor transmissions. Activated by the   energy harvesting (EH) from multiple  WPT sources, %process, 
the %multiple 
sensors transmit short packets to a destination with %via 
finite blocklength (FBL) codes. % (FBL).
This work  \emph{for the first time}  characterizes the FBL reliability for such multi-source WPT enabled network and accordingly provides %accordingly 
reliability-oriented resource allocation designs,  while a practical nonlinear EH model (including the effects of mutual interference among multiple RF signals) is considered. 
%To accurately describe the practical energy acquisition process, multiple RF sources based nonlinear EH model is utilized to formulate the resource allocation problems.
On the one hand,  for the scenario with a fixed frame structure, 
%we first characterize the 
we aim to maximize the FBL reliability via 
optimally allocating  the transmit power 
among the multiple %multiple %at different 
 WPT sources. %under scenarios with fixed wireless information transfer (WIT) frame structure. 
In particular, {we   investigate the relationship between the overall error probability and the transmit power  %of multiple RF signals 
of multiple WPT sources},  based on which  a power allocation problem is formulated. %^ under a total power constraint of the WPT sources.  
To solve the formulated non-convex problem, 
we first introduce auxiliary variables to make the problem analytically tractable, %transmit power of sensors as new variables to be optimized , 
% { (new optimization variables unclear. Introduce auxiliary variables, or alternatively treat ... as variables)}, 
based on %following 
which an iterative % successive convex approximation (SCA) based 
algorithm is proposed   while applying  successive convex approximation (SCA) technique to the non-convex components of the problem. %Consequently, an efficient sub-optimal solution is obtained. 
%{\color{cyan}On the other hand, in order to further improve system reliability, as well as to obtain a more general system design and thus enhance its scalability in a variety of  wireless scenarios,}
On the other hand,  we   extend our
%{the (our, show confidence)} 
design  into  a dynamic frame structure scenario, i.e., %in addition to the multiple source power levels, 
the blocklength allocated for WPT phase and short-packet transmission phase are adjustable,   which introduces more flexibility and new challenges to the system design.   %, which develops potential to further enhance system reliability.
%where two efficient  joint power and blocklength  allocation designs are accordingly provided. {
%(Q: Do we really need to mention alternating optimization method? A: The simulation figures are very small and more figures can be considered. So, no worries about the paper length. And almost no simulations about alternating optimization, right?)} %where resource scheduling is considered among both sources and users.
%We prove that the overall error probability is jointly convex in the power levels at sources and the blocklengths allocated to users under linear EH model. 
%{an alternating (optimization) based resource allocation design}, % is proposed, in which  independent power allocation and blocklength allocation are alternately performed  till the result converges. 
% As the 
% %{a (the, `a' means that the second one is unimportant)} 
% second method, 
{In particular, we provide a joint power and blocklength allocation design to minimize the   overall error probability under the total power and blocklength constraints.} 
%{\color{cyan}Nevertheless, 
%we encounter a much more challenging 
An optimization problem with high-dimension  variables is formulated, which suffers from the complex and non-convex relationship among system reliability, multiple source power and blocklength.
To tackle the difficulties, auxiliary variables introduction, multiple variable substitutions along with SCA technique utilization are exploited to reformulate and efficiently solve the problem.
% a high-quality solution with low complexity.
%and are proposed to minimize the overall error probability. 
% In the alternating based method,  
% While in joint optimization based method, power and blocklength are simultaneously optimized through variable substitution and SCA algorithm.
%, we  is applied to convert the original problem into local convex problems while auxiliary variables are introduced to facilitate the reformulation
Finally, through numerical results, we validate our analytical model %, 
and
evaluate the system performance, where a set of guidelines for  practical system design are concluded.
\end{abstract}

 \vspace{-.4cm}
\begin{IEEEkeywords}
  \vspace{-.39cm}
Energy harvesting, finite blocklength, resource allocation, short-packet transmission,  %wireless power transfer,
multi-source~WPT %{(use mWPT as short for multi-source WPT?)}
\end{IEEEkeywords}
  \vspace{-.4cm}

%{\color{blue}Blue: Changed. To be verified.}

%{Red: To be corrected or to be checked.}

%{\color{yellow}Yellow: To be deleted.}

\vspace{-.2cm}
\section{Introduction}
\vspace{-.1cm}
%1. {\color{blue}Communication Requirements (IoT/special scenario/EE...) $\rightarrow$ Batteries disadvantage$\rightarrow$ WPT a promising solution. }
Rapid developments of the Internet of things (IoT) are expected to enable the data transmissions for %of 
massive types of devices~\cite{Enabling2021Guo}. 
The upcoming industrial wave will present a great deal of difficulty in supplying power to a tremendous number of IoT devices~\cite{Wirelessly2020Wang}. Conventional power supplement  methods (e.g., cabling, batteries) are facing challenges due to 
%{the limitations of their costs (
their limitations in the costs, capacity and lifespan. 
In addition, laying out wired connections or replacing batteries %{may be inconvenient (
may lead to a huge inconvenience especially for the massive-device IoT network and even a high risk for certain applications e.g., metallurgical industry. 
In recent years, wireless power transfer (WPT) technology has been proposed and verified 
%(proved or verified?)} 
as a practical and promising solution~\cite{performace2018Morsi,Wireless2018zhang}. %~\cite{performace2018Morsi,Wireless2018zhang,wireless2014chen,Multi2014liu}.
WPT allows the power transmission 
without any physical connections or exposed contacts between sources and electrical devices,
%to address the issue. 
{In fact, WPT has been utilized in numerous applications with low power transformation, e.g., electric vehicles~\cite{Research2021Xiong}, radio frequency identification (RFID) tags~\cite{Numerical2018Buffi}, biomedical implant equipment~\cite{A2021Hassan} and other fields. 
Considering its advantages of safety, reliability and flexibility~\cite{Recycling2021Sun,A2022Hou}, WPT introduces a new approach for energy acquisition and alleviates the over-dependence on batteries, which is 
% In particular, %Enabled by WPT, the power transmission can be performed} 
%  which is 
 highly recommended in the scenarios, %necessary in certain scenarios, 
e.g., networks with mobile sensors and inaccessible remote systems.} %, etc. 
% {Considering the advantages of safety, reliability, and flexibility, WPT has already been widely utilized in biomedical implant equipment [9], [10], electric vehicles [11], [12], and other fields. }
% Nevertheless, it is worth mentioning that the received radio frequency (RF) energy from WPT sources cannot be directly utilized by devices. During the energy harvesting (EH) process, the RF signals received by the device can be converted into direct current (DC) signals through a rectifier circuit~\cite{fundamentals2019clerckx}. Afterwards, the applicable DC signals can be utilized for EH,  which introduces a new approach for energy acquisition and alleviates the over-dependence on batteries. 
 
% During the energy harvesting (EH) process, the 
% For energy acquisition, 
% {Via (not via. Conversion is a necessary part for EH, not the results after EH.)} energy harvesting (EH) process, the received radio frequency (RF) signals at IoT devices can be converted to the direct current (DC) signals through a rectifier circuit~\cite{fundamentals2019clerckx}, {(Seems uncompleted. Like, DC signal is used for EH.)} which introduces a new approach for energy acquisition and alleviates the over-dependence on batteries. 

% {(Transition sentence missing. Like words, WPT can also help communication.)} %% this was only an example for showing transition.

Moreover, integrating WPT technology into wireless network 
also assists to %the 
support the potentially massive unsourced %for  unsourced 
wireless devices. 
%WPT can also help communication.
So far, the WPT technology has been implemented in a variety of wireless scenarios, including cellular networks~\cite{Joint2021zhu}, mobile edge computing (MEC) systems~\cite{Computation2018Zhou,Energy2021Malik,Wireless2021Li}, unmanned aerial vehicle (UAV)-assisted communications~\cite{Joint2022Yuan,Sustainable2021Hu,UAV2021Zheng}, multiple-input multiple-output (MIMO) systems~\cite{Optimal2022Shanin,Wireless2016Zhu,Wireless2019Kamga}, etc.  
% {Considering the limited energy conversion efficiency during EH process, system energy efficiency may be unsatisfying. (unclear)} Regarding this, numerous efforts have been made to enhance the performance of WPT enabled networks. 
Considering the limited conversion efficiency during EH process, the limited applicable electrical energy may result in unsatisfying system reliability, efficiency, throughput, etc. Regarding this, numerous efforts have been made to enhance the performance of WPT enabled networks.
% { references:minimize/maximize the latency/reliability/throughput via time/power/blocklength allocation... }
For instance, authors in~\cite{On2019Waqar} studied a WPT assisted relaying network and proposed a throughput maximization design through beamforming vector and power splitting (PS) factor optimization.  
%{maximized {throughput (tight/direct connection between sentences is missing, to be reorganized)} of an amplify-and-forward-based relay network through beamforming vector and power splitting (PS) factor design. 
In~\cite{Energy-Efficient2017Chang}, a resource allocation scheme, including power allocation and time division protocol, was proposed to improve the energy efficiency in a wireless powered massive MIMO system.
% To improve the energy efficiency in a WPT powered massive MIMO system, a resource allocation scheme {(beamforming design, antenna selection, power allocation and time division protocol) (too much. Perhaps only mention the last two terms.)} is proposed in~\cite{Energy-Efficient2017Chang}.
Authors in~\cite{Non2022Wu} investigated non-orthogonal multiple access (NOMA) assisted federated learning activated by WPT and  a joint optimization design is proposed to minimize a system-wise cost. %{(Nice one)}
%{However (Nevertheless, two however is too much.)},
Nevertheless, all of the aforementioned results are conducted under the assumption of a linear EH process, i.e. the harvested power is proportional and linearly related to the received power, which is likely %{\color{cyan}sufficiently} 
inaccurate in practice. 
In realistic scenarios, as a consequence of the component nonlinearity in rectifier circuits, the output DC power is a complex nonlinear function of the received RF power.
% and the energy conversion efficiency {increases first and then decreases with the increasing received RF power.(a little bit wordy)} {(The important statement is lack of references.)} 
Therefore, for practical implement, nonlinear EH model is essential, especially in wireless power communication networks, where the power  has significant effects on the system performance. 
%('use' too oral, not elegant.)}
% to ensure the actual performance of the system, 
% analyze the practical system performance and provdee  practical EH model is 
% Based on the practical nonlinear EH model, 
%{(No smooth connection.) 
In our previous works, WPT enabled low-latency relaying networks~\cite{simultaneous2021Yuan} and  multi-user cooperative networks~\cite{SWIPT2021Guo} have been studied while considering this nonlinearity. % nonlinear EH model.

So far, numerous system designs have been proposed based on the nonlinear EH model, but basically conducted in the single WPT source scenarios~\cite{Wave2022Feng,Wireless2018Kang,Optimal2022Xu,energy2016wu}. 
%{(lack of smooth transition and connection.) 
In practical applications, due to pathloss and limited RF-DC conversion efficiency~\cite{Weighted2020Wu}, the WPT efficiency is usually around a low  level, which makes it  challenging to enable reliable and effective     communications %especially in networks with free positioning devices, %{system performance requirements (too vague) 
with a single WPT source~\cite{Optimizing2018Arakawa}. 
% Furthermore, single source enabled WPT cannot sufficiently satisfy the requirements of networks with multiple users or free positioning devices, where a large range of power transmission is required. 
{For this issue, numerous efforts have been made for WPT performance enhancement, including backscatter communications~\cite{Ye2022Mutualistic}, waveform design~\cite{Refined2021Abeywickrama}, deployment of directional antenna~\cite{joint2021yuan}.}
Besides, driving WPT with multiple sources is deemed to be an effective solution, which enables higher WPT efficiency and energy supply ceiling, thus accordingly achieves further system performance enhancement.
% Early in~\cite{convex2014lang}, the authors discussed the potential of enhancing the WPT efficiency via applying  multiple transmitters.
On the one hand, single source has limited WPT capacity, i.e., the transmit power from %available by 
a single source 
is generally upper-bounded by the hardware. %limited. 
Consequently, the received power at devices is likely to be rather lower and thus %, which 
cannot ensure efficient WPT. 
By contrast, collaborative utilization of the multiple WPT sources can potentially promote EH conversion efficiency~\cite{convex2014lang} and correspondingly enhance the overall WPT efficiency. 
%based on the nonlinear EH model {[?]}, {the growth of charged power created from each new unit of received power slows down progressively (unclear about the meaning)}, indicating that relying on a single WPT source may result in resource waste.
% By contrast, multiple sources is capable of making more efficient use of the limited resource via reasonable power allocation, i.e., all sources can achieve relatively high energy conversion efficiencies in the nonlinear model, and correspondingly improving {the energy conversion efficiency (reused words)} of the whole system.
On the other hand, channel %signal 
fading is a non-negligible factor in practical wireless %and varies between 
scenarios.
%channels. 
Compared with a single source, channel diversity introduced by multiple sources is capable of compensating deep fading. Namely, with multiple WPT sources, in case of deep fading at certain channels, the devices can benefit from %can select 
one or more relatively superior channels (sources) for high-efficiency energy transmission.
This diversity against fading can also improve the system robustness for wireless powered communication. 
%networksrobustness for location-independent systems.
%{(A little bit confusing. How about ``the system robustness for wireless powered communication networks.'')} % performance .
%robustness to system performance with respect to devices locations.
%On the other hand, the fading of RF signals propagating in space cannot be ignored and the pathloss increases with distance. Single WPT source can hardly guarantee reliable WIT for the sensor far away from it, especially in a multi-sensor network.
%, which announces higher requirement of source power for WPT. 
%Larger source power can indeed contribute to larger charged power, however, as well as leads to relatively lower energy conversion efficiency based on the nonlinear EH model.
%By contrast, multiple WPT sources can make up for this deficiency since a significantly larger area is covered {(Diversity against fading?)}, which introduces robustness to the reliability performance with respect to sensors locations.
%{(1. The whole paragraph has no references. 2. Nice attempt for analysing the necessity of multi-source WPT. Just not clear enough. Maybe we can improve it by clarifying (rough words) 2.1 single source with limited WPT capability (limited maximum transmit power) cannot ensure high efficient WPT. Multiple WPT sources can potentially promote conversion efficiency and correspondingly the overall WPT efficiency. 2.2 Diversity against fading / compensating deep fading. 3. Sentences need to be reclarified with tighter logic.)}

%%%%%%%%%%%%%%%%%%%%%%%%%%%%%%%
% 近义词不是同义词 每次用新词替换最好检查新单词原本的意思
%%%%%%%%%%%%%%%%%%%%%%%%%%%%%%%

Early in~\cite{convex2014lang},  
the authors %in ~\cite{convex2014lang} analyze 
discussed the potential of enhancing the WPT efficiency via applying  multiple transmitters.  
However, it is important to note that multiple sources will bring about a great deal of challenges to system designs.
In work~\cite{federated2022Hamde}, the performance of multi-source WPT  is studied, in which the sources operate  in an independent manner, i.e.,  a joint design among the WPT sources is ignored. 
Indeed, % fact, 
when multiple sources simultaneously transmit RF signals to users, the interference between the multiple RF signals will significantly affect the waveform of the received signal at devices and the interference has the potential to have either a positive or negative effect on WPT efficiency.
%{(? Lack of introduction other multiple-source works, which were indeed done in conference version.)} 
%{From the perspective of hardware, diverse signal power levels from multiple sources will result in different nonlinearity during EH process. (not convincing)} 
%Moreover, 
%{(abruptly mentioning mutual independent, may make readers confusing. No smooth organization.)} {when multiple sources transmit  mutual independent RF signals to user simultaneously, the interference between the multiple received RF signals cannot be ignored (this is not particularly for independent signals)}, which will significantly affect the waveform of the received signal at devices. 
% In work~\cite{federated2022Hamde}, the performance of multi-source WPT is studied, but the interaction between the WPT sources is ignored. 
In other words, %In addition, 
inappropriate resource allocation design may even decay the WPT performance, %
%{competition between multiple sources is inevitable, especially under a finite amount of resources (competition ?) (Does it mean inappropriate resource allocation design may even decay the WPT performance?)}
which necessitates advanced resource allocation strategies for system performance enhancement. 
%resource allocation strategy is crucial to maxi for and 
% Advanced resource allocation designs are crucial in enhancing system performance.
% Multiple WPT sources can enhance energy conversion efficiency and accordingly improve system performance, but the characterization of charged power is hampered by formidable challenges, and the modeling of the problem becomes more complicated.
Nevertheless,  for multi-source WPT system, an analytical EH
model has been proposed recently 
%investigated~
in \cite{convexity2022yuan}.
However, to the best of our knowledge, fundamental characterizations on the overall performance  for  multi-source WPT enabled multi-user transmissions, and the corresponding resource allocation designs for system performance improvement 
are  still  missing in the literature.
Moreover, it is worthwhile to mention that the data packets generated from IoT devices are likely to be short, i.e., the data transmissions are perhaps
%{are usually (inconsistent with likely, are perhaps)}
operated via finite blocklength (FBL) codes. Hence, %{the above characterizations(?)} and designs are
the FBL impacts are   essential to be   taken into account in the performance analysis and system designs for such networks.
In this paper, we consider a multi-source wireless powered communication network, where practical nonlinear model is considered for multi-source enabled EH. We characterize the system reliability under the FBL regime and propose a power allocation design to minimize the overall error probability under fixed frame structure. 
To further improve the system reliability and explore more general system designs, we extend our work into scenarios with dynamic frame structure, under which a joint power and blocklength allocation design is provided.
Our main contributions   are listed as follows. %{(Contributions are more about the value of the work)} 
\begin{itemize}
    %\item {Scenario: Multi-source WPCN}
    \item {\bf{{FBL Reliability characterization in Multi-source WPT enabled network}}}: For the first time, we study a multi-source WPT enabled short packet transmission network, in which the interference introduced by multiple WPT signals is considered and a RF signal combining strategy is utilized during the energy conversion process. %Our  … proposed joint multi-source enabled network is capable of providing significant energy efficiency improvement and has potential to be extended to a variety of scenarios, e.g., providing better energy supply strategies for MEC systems, UAV communications and cellular networks and accordingly achieve better system performances, including system reliability, delay, throughput, etc.
    {Different from previous works regarding convexity characterization of multi-source WPT, we characterize the system overall FBL reliability particularly for multi-sensor short packet transmissions, which is jointly affected by multi-source power, WPT blocklength as well as multiple WIT blocklength.}
    % , We characterize the relationship among the overall error probability, multi-source power, WPT blocklength as well as multiple WIT blocklength for multi-sensor short packet transmissions. 
Without loss of model accuracy, %To accurately model the practical EH circuits, 
a realistic nonlinear model for EH has been considered.
    \item
{\bf{Reliability-oriented resource allocation designs}}:
%{
%Following the characterization and 
Considering both cases of fixed frame structure and dynamic frame structure, we focus on the power allocation design and the joint power and blocklength allocation design targeting minimizing the overall error probability.
Due to the nonlinearity of  EH process and the sophisticated FBL reliability model, %is not convex in multiple source power, which makes 
both aforementioned problems are non-convex.
%which makes the optimization problem non-convex and cannot be directly solved through convex optimization tools. 
To cope with it, {we decouple the complex relationship between FBL reliability in WIT phase and the power allocation decision during WPT phase and  introduce %dependent variables (decided by multiple source power) as 
auxiliary power variables representing the adopted transmit power for short packet transmissions, while variable substitution has been proposed particularly for the joint resource allocation. }
The above approaches have made the reformulated problems to be more tractable, such that the successive convex approximation (SCA) can be implemented for obtaining suboptimal solutions for the resource allocation problem via iterative algorithms.  %, the non-convex problems are reformulated to be a high tractable ones. % for problem reformulation 
%with respect to which the error probability is convex.
%Through variable substitution, the harvested power is proved jointly convex in the substituted multiple source power.
%This result, along with the inequality constraint between substituted multi-source power and auxiliary variables, facilitates the further utilization of successive convex approximation (SCA) algorithm, by which the original non-convex constraint with respect to nonlinear EH process can be converted to a convex one and a sub-optimal solution  to the reformulated problem can be obtained based on iterative algorithm.
{Such   methodology, especially the introduced approach of non-convex problem transformation %for problem tractability enhancement
% Such a design methodology (including problem transformation tricks)  for enhancing the problem tractability %the non-convex problem reformulation 
 can be clearly extended to }multifarious multi-source enabled %and the joint convexity with respect to charged power and substituted  multi-source can be applied in  multi-source 
wireless powered networks and largely facilitate the corresponding resource allocation designs. %other types of resource allocation designs in. % under a variety of scenarios.

    \item
{\bf{Simulative validation and evaluation}}:
Via %Monte Carlo 
simulations, we confirm the analytical model and
evaluate system performance under the proposed     designs. 
%In particular, we identify that larger total source power, longer blocklength, smaller packet sizes as well as more  WPT sources  lead to a higher system reliability. %Moreover, o
In particular, the proposed design has shown to significantly outperform the benchmark and such reliability enhancement is expanded while more resources (power and blocklength) are provided in the system. In addition, the high  reliability robustness %to reliability 
with respect to sensor locations, which results from our introduced multiple WPT sources, has also been illustrated. % multi-source design introduces.
\end{itemize} 

The rest of this paper is organized as follows: In Section~\ref{sec:preliminaries}, we describe the multi-source WPT   system and review the FBL communication   model along with the nonlinear EH model. 
In Section~\ref{sec:power allocation}, we propose the power allocation design maximizing the characterized   overall reliability, %under the fixed frame structure,% considering both linear and nonlinear EH model. 
%{Following that (following what?)}, the work is
which is further extended to joint power and blocklength allocation design %with flexible frame structure 
in Section~\ref{sec:Power and Blocklength Joint Allocation}. 
Finally, we provide numerical results in Section~\ref{sec:numerical results} and conclude this work in Section~\ref{sec:Conclusion}.

% \noindent\emph{Notation:} Boldface letters represent vectors. $\Vert \cdot \Vert$, $P_r\{\cdot\}$ and $E_x[\cdot]$ denote the Euclidean norm, probability and statistical expectation of $x$, respectively. $\dot f(x)$ and denotes the first order derivative.

\vspace{-.25cm}
\section{Preliminaries}
\label{sec:preliminaries}
\vspace{-.22cm}
% In this section, we first present the system model, % the multi-source WPT enabled multi-sensor communication network. 
% and subsequently review the FBL communication performance model  and nonlinear EH model.
%(just repeated in the previous paragraph)

%\vspace{-.5cm}
\subsection{System Description}
\begin{figure}[t]
\centering
\includegraphics[width= 0.63\textwidth, trim=2 141 10 169]{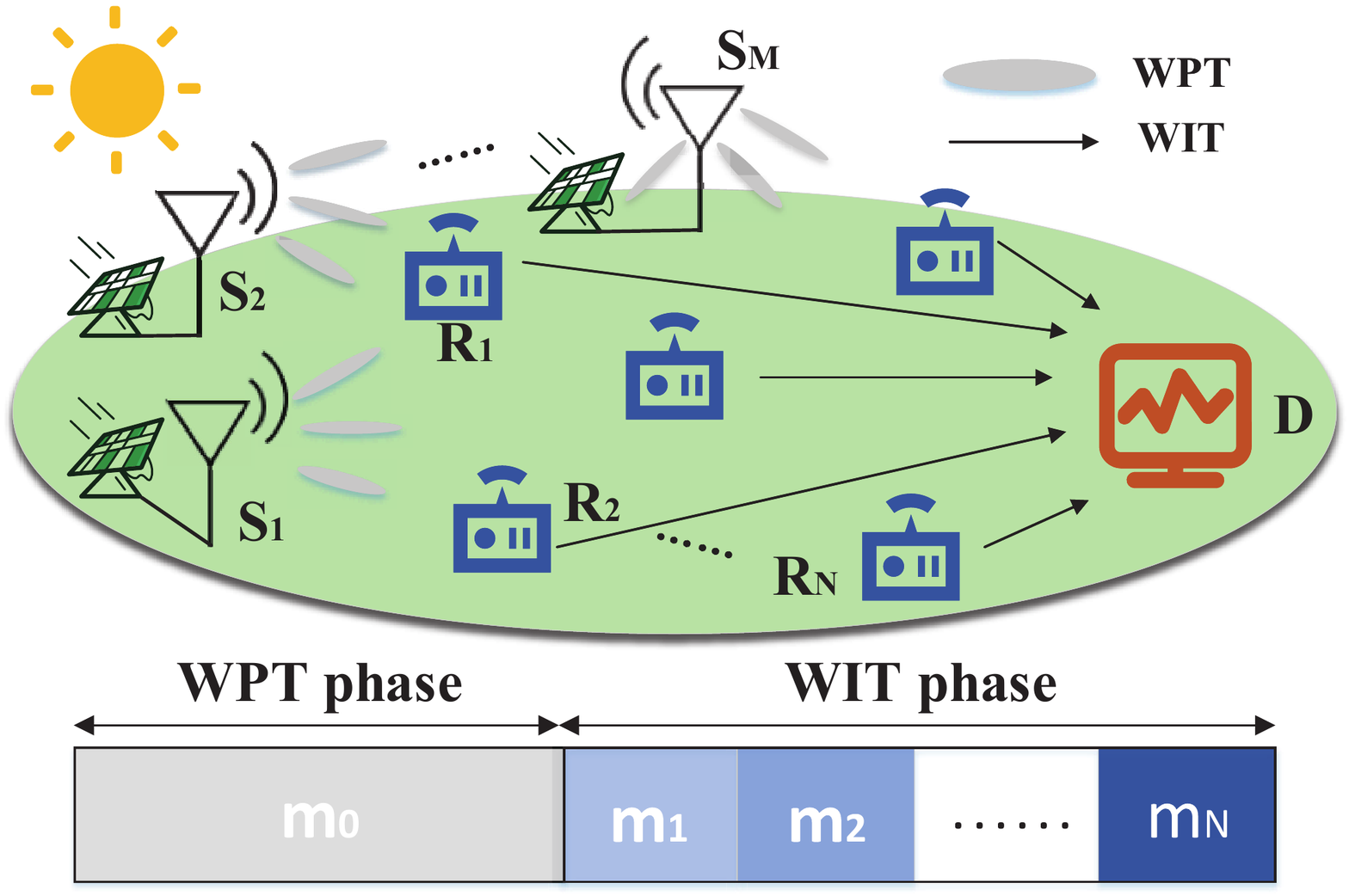}
\DeclareGraphicsExtensions.
\caption{System model and frame structure.}
\label{system model}
\vspace{-.95cm}
\end{figure}

\vspace{-.12cm}
We study a WPT enabled short packet communication network with $M$ WPT \underline{\bf{s}}ources (S), $N$ EH senso\underline{\bf{r}}s~{(R)} and a  \underline{\bf{d}}estination (D) as shown in Fig~\ref{system model}. 
We assume that these WPT sources  $\{ \rm{S}_j\}$, $j\in\{1,...,M\}$ have buffered   sufficient energy  (for performing WPT) from surrounding environment, e.g., via wires connection to remote photovoltaics and subsequently converting solar energy into electricity for local storage. % jointly perform WPT.
 %
 % for activating short packet transmission. 
Each sensor $\{{\rm R}_i\}$, $i\in\{1,...,N\}$ has a short packet with size of  $k_i$ bits needing to be transmitted to the destination $\rm{D}$, where a wireless information transmission (WIT)  process can be performed based on the energy harvested from the received RF signal (transmitted from the WPT sources). 
  % Owing to the   RF EH unit equipped at each sensor,  energy harvested from the RF signal power for wireless information transmission (WIT) to the destination.
For instance, a possible scenario would be  %{early warning (strange word, a specific concept?)} 
state monitoring system in volcano (or cave/mine) where numerous sensors equipped with RF EH unit are distributed  and a control center can provide early warning of an approaching eruption utilizing the monitored environmental information, which is measured and transmitted from multiple sensors.
In such  high-temperature (or temperature-sensitive) situation,  batteries may  not be the preferred energy supply for the senors, and WPT technology becomes a promising way to continuously and safely enable these sensors, 
%Due to the combustible nature of batteries in high-temperature situations, battery-powered data transmission is forbidden. {Thereby, WPT technology is applied (too plain)}, which can transfer energy in a timely and secure manner. 
while the WPT source can be recharged via   a wired line connected {to the} remote   nature EH device, e.g., solar  panels. 
%On the surface of the volcano, a variety of sources that can capture and store energy from nature are scattered. 
%All the sensors are equipped with RF energy harvesters, i.e, the received RF signal power from WPT sources can be harvested and subsequently the WIT process can be conducted via consuming the harvested energy.

The entire process is divided into two phases, i.e., a WPT phase and a WIT phase. During the WPT phase, instead of a single source, multiple WPT sources are utilized as energy suppliers in attempt to achieve a relatively higher energy efficiency.
In particular, $M$ sources transmit mutually independent RF signals to $N$ sensors in a broadcasting manner, with possibly different transmit power which are denoted by $\{P_j\},\forall j=1,\cdots,M$.
{The blocklength (i.e., the amount of symbols) of the WPT phase is denoted by $m_0$, while the time duration of a single symbol  is denoted by $T_s$. Consequently, the time length for WPT is  $m_0 T_s$. }
% The time length of the WPT phase is denoted by $m_0 T_s$, where $m_0$ denotes the length in symbol and $T_s$ is the time length of a single symbol.
Since sufficient energy has been buffered from the surroundings by sources, a significant quantity of transmit power can be generated for WPT. This, coupled with the generally small packet size and correspondingly relatively little energy required for reliable WIT, results in a relatively short time length for performing WPT. Therefore, it is reasonable to consider the channel gains within WPT phase as constant. The channel gain between ${\rm{S}}_j$ and ${\rm{R}}_i$ is denoted by~$z_{i,j}$.
After receiving the independent RF signals from multiple WPT sources, the EH unit of each sensor converts them to DC signals and performs EH to supply sufficient energy for the forthcoming short packet transmission. We assume that the EH units' EH capability is stable, i.e., the charged power at ${\rm{R}}_i$ is a constant and denoted by $Q_{\rm{ch},i},\forall i=1,\cdots,N$, such that the corresponding harvested energy during the whole WPT phase can be written as $Q_{\rm{ch},i}m_0 T_s$.

During the WIT phase, $N$ short packets with sizes of $k_i,\forall i=1,\cdots,N$ bits are transmitted to D from $N$ sensors by consuming the harvested energy. The WIT process is performed in a  time-division duplex (TDD) manner, i.e., $N$ packets are sequentially transmitted  to D, % in a certain order, 
with which the co-channel interference introduced by multiple sensors can be %directly 
avoided.
To ensure timely and {accurate transmission of}  the sensed data packets, low latency and high reliability WIT is required, which, coupled with the generally small data amount of the packets generated by IoT sensors, {indicates that} FBL codes have to be utilized for WIT. In the FBL regime, the decoding error probability cannot be ignored and the error probability of the transmission of each packet should be lower than 0.1 for guaranteeing ultra reliable communication, i.e., $\varepsilon_i\leq 0.1,\forall i=1,\cdots,N$ holds. 
Moreover, the entire duration of the WIT phase in symbol is restricted with $\sum_{i=1}^N m_i \leq m_{\rm{total}}$ , where $m_i,\forall i=1,\cdots,N$ denotes the blocklength of the packet transmission from ${\rm{R}}_i$ to D and $m_{\rm{total}}$ specifies the maximum total amount of symbols permitted during the WIT phase. {Note that   each  symbol has a fixed  length of   $T_s$. Hence, the blocklength restriction can be formulated as a time   constraint that ensures the low-latency WIT, i.e., $\sum_{i=1}^N m_i T_s \leq m_{\rm{total}} T_s$.}
Since the channels are assumed to be independent and experience quasi-static fading and the blocklength of each packet transmission is relatively short, the channel gain is reasonably considered as a constant during one block and varies independently to the next, i.e., the channel gain between ${\rm{R}}_i$ to D can be denoted by $\bar{z}_{2,i},\forall i=1,\cdots,N$. The noise power level is given as $\sigma^2$. 

In addition, %it is critical to 
note that each sensor will utilize all the harvested energy for short packet transmission (to maximize the utilization of the energy) and accordingly guarantee reliable transmission.
In this regard, the transmit power of ${\rm{R}}_i$ is jointly restricted by the harvested energy during  the WPT phase $Q_{\rm{ch},i}m_0 T_s$ and the time length of its packet transmission $m_i T_s$, i.e., $\bar{P}_i=\frac{Q_{\rm{ch},i}m_0 T_s}{m_i T_s}$, where $\bar{P}_i$ denotes the transmit power of ${\rm{R}}_i$ for the short packet transmission in the WIT phase. 

As discussed above, multiple sources are responsible for providing power supplement to multiple sensors, hence ensuring reliable short packet transmissions.
However, the practical energy conversion efficiency during the EH process is quite limited,  necessitating   characterization and optimal designs  on the achievable reliability performance of the considered network. % that the network is capable of supporting.

\vspace{-.4cm}
\subsection{FBL Transmission   Model}
\vspace{-.1cm}
%In our work, %for the short packet transmissions carried via FBL codes, we exploit   Polyanskiy's FBL model to 
%the FBL reliability is utilized as the indicator to %resource utilization efficiency and the topic of 
%how to allocate power among multiple sources under the total power constraint appears necessary and alluring.
%In addition, blocklength is another important factor that will affect the FBL reliability, which is also necessary to be considered.

%In comparison to the Shannon   capacity bound, the FBL model is more accurate when the blocklength is finite/short. 
The data transmissions in the WIT phase are assumed to be carried via FBL codes.
According to~\cite{channel2010polyanskiy,Cross2018She}, with error probability $\varepsilon$, signal-to-noise ratio (SNR)~$\gamma$, and blocklength $m$,  the coding rate  is given by %shown to have the following asymptotic expression:
 \vspace{-.2cm}
\begin{equation} \vspace{-.2cm}
r \approx {\cal C} \left( {\gamma} \right) - \sqrt { {\cal V}\left( {\gamma} \right)/m} {Q^{ - 1}}\left( \varepsilon  \right),
\end{equation} 
where  ${\cal C}\left( {\gamma} \right)\!=\!\log \left( {1 \!+\! {\gamma}} \right)$ and  ${\cal V}\left( {\gamma} \right)\!=\!\left(1 \!-\! \frac{1}{{{\left( {1 \!+\!\gamma} \right)}^2}}\right) {\log _2^2}e$ are  the Shannon capacity and the channel dispersion,  and $Q^{ - 1}(\cdot)$ is the reverse of
$Q\left( x \right)\! =\! \int\nolimits_x^\infty  {\frac{1}{{\sqrt {2\pi } }}} {e^{ - {t^2}/2}}dt$, which
represents the Gaussian $Q$-function. 
In the other way round, with the coding rate $r$,  blocklength $m$ and  SNR $\gamma$, the error probability can be expressed as 
  \vspace{-.3cm}
\begin{equation}  \vspace{-.2cm}
\varepsilon  ={\cal P}\left( \gamma, r,m \right) \approx Q\left( {\frac{{{\cal C}\left(\gamma \right) -  r}}{{\sqrt {{\cal V} \left( {\gamma} \right)/m} }}} \right).
\label{qfunc}
\end{equation}

\vspace{-.5cm}
 \subsection{Nonlinear Energy Harvesting Model} 
 \vspace{-.1cm}
During the EH process, the received RF signal can be converted to a DC signal for further EH.  
{So far, {several} nonlinear EH models have been established to depict the EH process, such as piece-wise linear based nonlinear EH model~\cite{Practical2015Boshkovska} and curve fitting based nonlinear EH model~\cite{Energy2019Shi}. Both of them are stochasticmodels, which are decided by the measurement data from practical EH circuits and require another experiment when changes take place in the circuit.
In~\cite{nonlinear convex} an analytical EH model based on practical EH circuits components is investigated, in which} the output DC of the EH unite is represented  by a nonlinear implicit function with respect to the received RF power and the energy conversion efficiency is particularly poor with low or high input power  due to the turn-on voltage and the reverse breakdown of the diode(s) used in the rectifier circuit.
{However, all the aforementioned nonlinear models only consider a single RF signal for EH.} When it comes to multi-source WPT, multiple RF signals are transmitted simultaneously and {the harvested energy is not the sum of harvested power over each single RF signal.}
% the characterization between the harvested energy and multiple received RF power  is significantly more complex, which is not a straightforward accumulation of harvested energy obtained from a single RF signal.
% For one thing, different RF power levels will result in various nonlinearity during the EH process. 
% For another thing, due to the simultaneous transmission of several RF signals to the user, mutual interference between the independent received RF signals cannot be ignored, which makes multi-source WPT enabled communication considerably more challenging than that with a single WPT source. 
Recently, an analytical EH model for a multi-source WPT system is investigated.
According to~\cite{convexity2022yuan}, the harvested DC power of  multi-source WPT with independent RF signals can be represented by 
  \vspace{-.1cm}
\begin{equation}  \vspace{-.1cm}
    P_{\rm{dc}} \triangleq F_{\rm{nl}}(\boldsymbol{Q}) \approx \left(\frac{1}{a} W_0(ae^{a I_{\rm{s}}} \varphi(\mathbf{Q}))-I_s\right)^2 \hat{R}_{\rm{L}},
    \label{Pch nonlinear}
\end{equation}
where  $\boldsymbol{Q}\triangleq \left({Q_1,\cdots,Q_M}\right)$ represents the set of  $M$ received RF signal powers. In addition, $W_0(\cdot)$ indicates the principle branch of Lambert $W$ function, which is the inverse relation of function  $f(x)=xe^x$. Constant $a=\frac{\hat{R}_{\rm{L}}}{n v_{\rm{t}}}$, where $\hat{R}_{\rm{L}}$, $n$ and $v_{\rm{t}}$ represent the   load resistance, the ideality factor and the thermal voltage respectively. $I_s$ is the reverse bias saturation current, which is ignored in linear EH model. Moreover, $\varphi(\mathbf{Q})$ is an explicit expression written as
 \vspace{-.1cm}
\begin{equation}  \vspace{-.1cm}
    \varphi(\mathbf{Q}) = I_{\rm{s}}+\sum_{i=1}^{n_0} \beta_i \sum_{\sum_{m\!=\!1}^M \!i_{\rm{m}}\!=i} \hat{\mathcal{C}}_{i,\{{i_{\rm{m}}}\}}~\prod_{m=1}^M Q_m^{i_m/2},
    \label{Pch nonlinear phi}
\end{equation}
where $n_0$ denotes the truncation order, which affects the accuracy for modelling the nonlinearity. Constant $\beta_i =\Bar{k_i}R_{\rm{ant}}^\frac{i}{2} $, where $\Bar{k_i}=\frac{I_s}{i!(n v_t)^i}$ and $R_{\rm{ant}}$ represents  the matched antenna impedance.{  In constant $\hat{\mathcal{C}}_{i,\{{i_{\rm{m}}}\}}=\frac{M!}{\prod_{m=1}^M i_m! }\prod_{m=1}^M \lambda_{m,i_m} $, $\lambda_{m,i_m}$ denotes the constant waveform factor for unit power. 
Expression $\sum_{m=1}^{M}i_m=i$ represents that any non-negative sequence $\{i_m\}$ must have a sum of $i$.
% Component $\sum_{m=1}^{M}i_m=i$ represents that the sum of any sequence of $M$ non-negative integers $i_m$ is equal to i.
}
{According to~\eqref{Pch nonlinear}, we can learn that the harvested power and the EH efficiency are affected by the received RF power.  More clearly, $\varphi(\mathbf{Q})$ in~\eqref{Pch nonlinear} depicts the }multiple RF combining process and is directly influenced by the multiple RF signal powers $\boldsymbol{Q}\triangleq \left({Q_1,\cdots,Q_M}\right)$ based on~\eqref{Pch nonlinear phi}. Expression $\prod_{m=1}^M Q_m^{i_m/2}$ in~\eqref{Pch nonlinear phi} indicates that the interference between multiple received RF signals   cannot be ignored and that such interference can make an either positive or negative effect on energy acquisition results, which makes multi-source WPT enabled communication considerably more challenging than that with a single WPT source.
Therefore, a reasonable power allocation among multiple WPT sources is significantly essential for high energy conversion efficiency realization.

% {(Maybe a chance for stating the impacts of multiple source on the model, as well as the analysis difficulty in the model.)}

So far, we have stated the system model, the FBL transmission model and the nonlinear EH model. Following them, resource allocation designs will be proposed in the following sections.

\vspace{-.25cm}
\section{Power Allocation Among Multiple WPT Sources}
\label{sec:power allocation}
\vspace{-.12cm}
% To go from the simple scenario, 
In this section, we focus on the power allocation design under the fixed frame structure, i.e., %multiple source power  are to be optimized while 
the lengths of the WPT phase and WIT are given. 
We first characterize the relationship among the FBL reliability and the multiple source power. Based on the characterization, an efficient power allocation design is proposed   to minimize the overall error probability under the total power constraint.

\vspace{-.3cm}
\subsection{FBL Reliability Characterization and  Problem Formulation}
\vspace{-.1cm}
\label{power allocation-problem formulation}
During the WPT phase, $M$ sources transmit RF signals with power  $\{P_j\},\forall j=1,\cdots,M$ to $N$ sensors simultaneously in a broadcasting manner. 
Accordingly, each sensor ${\rm{R}}_i,\forall i=1,\cdots,N$ will receive a set of RF signals with power    $\boldsymbol{Q_i}\triangleq(Q_{i,\text{1}},\cdots,Q_{i,j},\cdots,Q_{i,\text{M}})$, where $Q_{i,j}$ denotes the power exclusively obtained from ${\rm{S}}_j$ and is given as
 \vspace{-.2cm}
\begin{equation} \vspace{-.2cm}
   Q_{i,j}(P_j) = P_j z_{i,j}.
\end{equation}
 However, the received RF power cannot be directly consumed, i.e., RF signals are required to be converted to DC signals via {an} EH process and the corresponding charged power is~written~as
 %{(To be more rigorous, always use the function form Q(P) instead of Q, to avoid misunderstanding. Otherwise, charged power will be easily considered as constant by reviewers.)}
  \vspace{-.2cm}
\begin{equation} \vspace{-.2cm}
    Q_{\rm{ch,}i}(\boldsymbol{P}) = F_{\rm{nl}}\left(\boldsymbol{Q}_i(\boldsymbol{P}) \right),
    \label{Qch_original}
\end{equation}
where $F_{\rm{nl}}(\cdot)$ characterizes the relationship between the harvested DC power and the received multiple RF power  according to~\eqref{Pch nonlinear}.
With charged power $ Q_{\rm{ch,}i}$, the corresponding  harvested energy of ${\rm{R}}_i$  during WPT phase is given as $Q_{\rm{ch},i}m_0 T_s$, which will be completely consumed in the following WIT phase.
During the WIT phase, each sensor transfer{s} short packet to D with transmit power $\bar{P}_i, \forall i \in \{1,2,...,N\}$,
%and the transmit power of ${\rm{R}}_i$ is denoted by $\bar{P}_i$, 
which is limited by the harvested energy, i.e., 
  \vspace{-.1cm}
\begin{equation}  \vspace{-.1cm}
   \bar{P}_i( Q_{\rm{ch,}i}(\boldsymbol{P})) = \frac{Q_{\rm{ch},i}(\boldsymbol{P})m_0 T_s}{m_i T_s}=\frac{Q_{\rm{ch},i}(\boldsymbol{P})m_0 }{m_i}, \forall i \in \{1,2,...,N\},
   \label{3.1 barP}
\end{equation}
{where $m_0$ and $m_i$ are positive integers denoting the blocklengths respectively assigned for WPT and  WIT (from ${\rm{R}}_i$ to D). Notation $T_s$ denotes the time length of a single symbol}.
As a result, the SNR of the transmission from ${\rm{R}}_i$ to D can be written as 
  \vspace{-.1cm}
\begin{equation}  \vspace{-.1cm}
     \gamma_i(\bar{P}_i) = \frac{{\bar{P}_i} \bar{z}_{\rm2,i}}{\sigma_i^2}.
     \label{3.1 SNR}
\end{equation}
Note that, with given channel gain $\bar{z}_{\rm2,i}$, noise power $\sigma_i^2$, WPT blocklength $m_0$ and WIT blocklength $m_i$ (under fixed frame structure), the sensor transmit power ($\bar{P}_i$) as well as the SNR ($\gamma_i$) are directly and completely decided by the charged power ($Q_{\rm{ch},i}$) and both of which are complex nonlinear functions with respect to multiple source power  $\{P_j\},\forall j=1,\cdots,M$ based on~\eqref{Qch_original}, ~\eqref{3.1 barP} and~\eqref{3.1 SNR}.
In order to guarantee reliable transmission, we assume the network should guarantee $\gamma_i\geq1$, which implies respectively direct and indirect limitations on $\bar{P}_i$ and $\{P_j\},\forall j=1,\cdots,M$.

% {require a transition sentence?}

According to~(\ref{qfunc}), the FBL error probability of the transmission from ${\rm{R}}_i$ to D can be written as $\varepsilon_i  ={\cal P}\left( \gamma_i, r_i,m_i\right)$, where $r_i=\frac{k_i}{m_i}$. Clearly, with given packet size $k_i$ and WIT blocklength~$m_i$,~$\varepsilon_i$ is fully influenced by $\gamma_i$. Combined with~\eqref{3.1 SNR}, the transmission error probability from ${\rm{R}}_i$ to D can be written as %{(The same issue. Error can be denoted in a function form, and can be clearer.)}
  \vspace{-.1cm}
\begin{equation}  \vspace{-.1cm}
    %\varepsilon_i(\bar{P}_i) ={\cal P}\left(\gamma_i(\bar{P}_i)\right) .\\
        \varepsilon_i(\bar{P}_i)  ={\cal P}\left(\gamma_i(\bar{P}_i), r_i,m_i \right) .
    \label{3.1 error i}
\end{equation}
%{($r_i$ is not defined yet.)}
Since all the packets are expected to be transferred with ultra reliability, the overall error probability is applied as the performance indicator for system investigation.
%\begin{equation}
    $\varepsilon_{\text O} = 1-\prod_{i=1}^N (1-\varepsilon_i)$.
%\end{equation}
As mentioned before, the error probability of transmission of each packet must be lower than 0.1 to guarantee reliable transmission, i.e.,  $\varepsilon_i\leq 0.1, \forall i \in \{1,2,...,N\}$. Hence, $\varepsilon_i+\varepsilon_{i'} \gg \varepsilon_i \varepsilon_{i'}$, $\forall i,{i'} \in \{1,2,...,N\}$ holds, i.e., the high-order term $\varepsilon_i \varepsilon_{i'}$ is negligible in comparison to $\varepsilon_i+\varepsilon_{i'}$. As a result,
%Via ignoring the high-order term,
the overall error probability can be tightly approximated to
  \vspace{-.1cm}
\begin{equation}  \vspace{-.1cm}
    \varepsilon_{\text O}(\varepsilon_i) = 1-\prod_{i=1}^N (1-\varepsilon_i)   \approx \sum_{i=1}^N \varepsilon_i.
    \label{sum errors}
\end{equation}
So far, the relationships among the overall error probability $\sum_{i=1}^N \varepsilon_i$, SNR $\boldsymbol{\gamma}$, sensor transmit power $\boldsymbol{\bar{P}}$ and source transmit power $\boldsymbol{P}$ have been characterized. 
Following the characterization, the power allocation problem %with the objective of 
minimizing the overall error probability is formulated as follow: 
%{(A common issue in the formulated problem. The definition for $\bar P$ is not constraint. Differentiate definitions from constraints. Keep that in mind.)}
% Recalling our problem, we aim to guarantee the system reliability by minimizing the overall error probability through power allocation under total transmit power constraint. The power allocation problem can be formulated as follows:
 \vspace{-.4cm}
\begin{subequations}
	\begin{eqnarray}
		\label{OP3.1:power_linear}
({\rm OP1}):&\!\!{\min\limits_{\{\boldsymbol{P}\}}}\, &\sum_{i=1}^N \varepsilon_i  \label{OP3.1obj:error}\\
    &\rm{s.t.:} &  \frac{{\bar{P}_i}(  Q_{\rm{ch,}i}(\boldsymbol{P})) \bar{z}_{\rm2,i}}{\sigma_i^2} \geq  \max\{1, 2^\frac{k_i}{m_i}-1\},\label{OP3.1con:SNR bound}\\
%	&\!\!& \bar{P}_i\leq y_{\rm{th}} ,\label{OP3.1con:y bound} \\
	&\!\!& \sum_{j=1}^M P_j \leq P_{\rm{total} },\label{OP3.1con:sum P}\\
		&\!\!& 0\leq P_j\leq P_{\rm max},\label{OP3.1con:upper P}
	\end{eqnarray}
\end{subequations}

%{(missing constraint: $0\leq P_j\leq P_{\rm max}$. Set up a proper $P_{\rm max}$ in simulation. Then the simulation figures do not need to be modified. Also in the following, update this constraint for all problems, including reformed and approximated.)}
 \vspace{-.2cm}
\noindent{where} objective function \eqref{OP3.1obj:error} presents the overall error probability, which has been tightly approximated as the sum of $N$ single link transmission error probabilities during the WIT phase accordingly to~\eqref{sum errors}.
% {Constraint~\eqref{OP3.1con:y value} announces the restriction on the transmit power of ${\rm{R}}_i$. (Here, $\bar P$ is a function in $P$, not a variable, right? If so, this is not a constraint. After correction, do not forget to correct all related statements.)}
%the relationship between the multiple transmit power at WPT sources  and the harvested power at EH sensors is depicted.
%\eqref{OP3.1con:y bound} announces the upper bound of the sensor transmit power.
In constraint~\eqref{OP3.1con:SNR bound}, term $\frac{{\bar{P}_i}(  Q_{\rm{ch,}i}(\boldsymbol{P})) \bar{z}_{\rm2,i}}{\sigma_i^2}$ represents the SNR of transmission from $\rm R_i$ to D. %based on~\eqref{3.1 SNR}. 
Constraint~\eqref{OP3.1con:SNR bound} specifies that the coding rate should be smaller than the Shannon capacity, i.e., $\frac{k_i}{m_i}\leq \log_2(1+\bar{P}_i \frac{ \bar{z}_{\rm2,i}}{\sigma_i^2})$ and the SNR should be larger than 1 to guarantee reliable transmission.
In~\eqref{OP3.1con:sum P}, total power consumption restriction is announced. Finally,~\eqref{OP3.1con:upper P} announces the upper bound of the transmit power for each WPT source.
By solving Problem (OP1), we can identify the optimal system reliability achievable under a given total power constraint and the corresponding optimal power allocation strategy.

However,   Problem (OP1) is not convex. 
{On the one hand, the objection function is not necessary convex, i.e., the convexity of error probability $\varepsilon_i$ to multiple source transmit power  $\boldsymbol{P}$ is unpredictable since $\varepsilon_i$ is directly affected by the transmit power $\bar{P}_i$, which is a complex nonlinear function respect to $\boldsymbol{P}$.  
On the other hand, the convexity of constraint~\eqref{OP3.1con:SNR bound} is also not guaranteed since term ${\bar{P}_i}(  Q_{\rm{ch,}i}(\boldsymbol{P}))$ is not concave.}
As a result, Problem (OP1) is non-convex in its current form, which cannot be efficiently solved by standard convex programming methods..

 \vspace{-.4cm}
\subsection{Problem Reformulation and SCA-based Solution }
 \vspace{-.1cm}
\label{sec1.3}
Based on prior analysis, original problem (OP1) is a complicated  non-convex optimization problem due to the nonlinear EH process. 
To address this issue, we first introduce auxiliary variables and subsequently reformulate the original problem via the SCA algorithm.

Though $\varepsilon_i$ is not necessary convex to $\boldsymbol{P}$, the convex relationship between $\varepsilon_i$ and SNR $\gamma_i$ has been demonstrated in~\cite{SWIPT2019hu}. 
Moreover, recalling~\eqref{3.1 barP} and~\eqref{3.1 SNR}, $\gamma_i$ is clearly and fully determined by $\bar{P}_i$ in a linear form and $\bar{P}_i$ is fully affected by multiple source power, which motivates us to introduce $\boldsymbol{\bar{P}}$ as a new set {of} variables (to be optimized). 
In other words, as a set of intermediate variables, $\boldsymbol{\bar{P}}$ can decouple the complex nonlinear relationship between the error probability $\varepsilon_i$ and WPT source power  $\boldsymbol{P}$ without modifying the original problem, which has the following two advantages: 
i. the equality function~\eqref{Qch_original} is then converted to an  inequality one, which enables the further application of SCA.
ii. original nonlinear constraint \eqref{OP3.1con:SNR bound} (with respect to $\boldsymbol{P}$) becomes linear (wtih respect to $\bar{P}$).
By this means, the reformulated problem is depicted as follows
 \vspace{-.3cm}
\begin{subequations}
	\begin{eqnarray}
		\label{OP2}
({\rm P1}):&\!\!{\min\limits_{\boldsymbol{P},\boldsymbol{\Bar{P}}}}\, &\sum_{i=1}^N \varepsilon_i  \label{3.3P1:objerror}\\
    &\rm{s.t.} &  \bar{P}_i \leq  F_{\rm{nl}}[\boldsymbol{Q}_i(\boldsymbol{P}) ]\frac{m_0}{m_i},\label{3.3P1con:barP}  \\
% 	&\!\!& \frac{{\bar{P}_i} \bar{z}_{\rm2,i}}{\sigma_i^2} \geq  \max\{1, 2^\frac{k_i}{m_i}-1 \} ,\label{3.3P1con:SNR constraints} \\
    	&\!\!& \frac{{\bar{P}_i} \bar{z}_{\rm2,i}}{\sigma_i^2} \geq  \max\{1, 2^\frac{k_i}{m_i}-1\},\label{OP3.3P1con:SNR bound}\\
	&\!\!&  ~\eqref{OP3.1con:sum P},  \eqref{OP3.1con:upper P}. \nonumber 
	\end{eqnarray}
\end{subequations}

 \vspace{-.4cm}
First, we take a look at~\eqref{3.3P1con:barP}, which includes a sophisticated nonlinear EH process. 
According to~\cite{convexity2022yuan}, $ Q_{\rm{ch,}i}$  is jointly convex in $\{\hat{Q}_{i,j}\}$ where $\hat{Q}_{i,j}=\frac{1}{Q_{i,j}}, j=1,\cdots, M$. 
 {\color{blue} }For ease of characterization, we define a set of new variables $\hat{\boldsymbol{P}}\triangleq \{ \hat P_\text{1},\hat P_\text{2},\cdots,\hat P_\text{M}\}$, where $\hat{P_j}=\frac{1}{P_j}, \forall{j}\in \{1,\cdots,M\}$.
 In particular,  $\hat{Q}_{i,j}=\hat{P_j}\frac{1}{z_{i,j} }$, where $z_{i,j}$  is a positive constant. 
Note that $ Q_{\rm{ch,}i}$ is jointly convex and apparently non-increasing in  $\hat{Q}_{i,j}$, and  $\hat{Q}_{i,j} $ is an affine function of $\hat{P_j}$. As a result, $Q_{\rm{ch,}i} = F_{\rm{nl}}[\boldsymbol{\hat Q}_i(\boldsymbol{\hat P}) ]$ is jointly convex in $\boldsymbol{\hat{P}}$ based on the convex property of the composite function. %{require details?about composite convexity } 
Therefore, constraint~\eqref{3.3P1con:barP}, \eqref{OP3.1con:sum P} and~\eqref{OP3.1con:upper P} can be respectively reformed as,
  \vspace{-.2cm}
 \begin{equation}  \vspace{-.2cm}
    \bar{P}_i \leq  F_{\rm{nl}}[\boldsymbol{Q}_i(\boldsymbol{\hat{P}}) ]\frac{m_0}{m_i},
     \label{3.3 hatPconstraint}
 \end{equation}
  \vspace{-.5cm}
 \begin{equation} \vspace{-.3cm}
    \sum_{j=1}^M \frac{1}{\hat{P_j}}\leq P_{\rm{total}},
     \label{constraint hatP sum}
 \end{equation}
  \vspace{-.3cm}
  \begin{equation} \vspace{-.3cm}
     %0\leq \frac{1}
  {\hat{P_j}} \geq \frac{1}{P_{\rm max}}. %(\text{reformed. Otherwise, this constraint is not convex.})
     \label{constraint hatP upper bound2}
 \end{equation}
However, constraint~\eqref{3.3 hatPconstraint} is not convex due to the non-concave  term $F_{\rm{nl}}[\boldsymbol{Q}_i(\boldsymbol{\hat{P}}) ]$. To address it, SCA is applied to convert the original non-convex constraint into a convex one.
In particular, a concave function $\hat{P}_{\rm{ch},i}$ is required, which satisfies $F_{\rm{nl}}[\boldsymbol{Q}_i(\boldsymbol{\hat{P}}) ]\geq \hat{P}_{\rm{ch},i}$. Since $Q_{\rm{ch,}i}=F_{\rm{nl}}[\boldsymbol{Q}_i(\boldsymbol{\hat{P}}) ]$ has been proved being joint convex in $\boldsymbol{\hat P}$, an inequality can be obtained based on the property of convex function
  \vspace{-.1cm}
 \begin{equation}  \vspace{-.1cm}
    Q_{\rm{ch,}i}=F_{\rm{nl}}[\boldsymbol{Q}_i(\boldsymbol{\hat{P}}) ]\geq \sum _{j=1}^M -A^{(\tau)}_{i,j}\hat{P_j}+B^{(\tau)}_i,
    \label{3.3 convert}
      \vspace{-.1cm}
\end{equation}  \vspace{-.1cm}
where  $A_{i,j}^{(\tau)}$ and $B_i^{(\tau)}$ are the positive constants defined as
  \vspace{-.1cm}
\begin{equation}  \vspace{-.1cm}
\begin{split}
 A_{i,j}^{(\tau)} & = \frac{\partial F_{\rm{nl}}[\boldsymbol{Q}_i(\boldsymbol{\hat{P}})]}{\partial \hat{P_j} }\mid_{\hat{P_j}=\hat{P_j}^{(\tau)}} ,
\end{split}
\label{A1}
\end{equation}
 \vspace{-.3cm}
\begin{equation} \vspace{-.3cm}
    B_j^{(\tau)}= F_{\rm{nl}}[\boldsymbol{Q}_i(\boldsymbol{\hat{P}})]\mid_{\hat{P_j}=\hat{P_j}^{(\tau)}}+ \sum_{j=1}^M A^{(\tau)}_{i,j}\hat{P_j},
\label{B1}
\end{equation}
and the equality in~\eqref{3.3 convert} holds at the local point $\boldsymbol{\hat{P}}^{(\tau)}$.
Therefore, the non-concave term  $F_{\rm{nl}}[\boldsymbol{Q}_i(\boldsymbol{P})]$ is approximated to a concave one $\sum _{j=1}^M -A^{(\tau)}_{i,j}\hat{P_j}+B^{(\tau)}_i$.
Consequently, the original non-convex problem is approximated into a local subproblem, which can
be solved iteratively until the stable point is achieved.
In the $\tau$-th iteration, the corresponding subproblem can be written as
  \vspace{-.1cm}
\begin{subequations}
	\begin{eqnarray}
		\label{OP3}
\!\!\!({\rm SP1}):&\!\!\!\!\!\!{\min\limits_{\boldsymbol{\hat{P}},\boldsymbol{\Bar{P}}}}\, &\sum_{i=1}^N \varepsilon_i  \label{2obj:varepsilon}\\
    &\rm{s.t.} &  \bar{P}_i \leq   \frac{m_0}{m_i}\!\left(\sum _{j=1}^M -A^{(\tau)}_{i,j}\hat{P_j}+B^{(\tau)}_i\!\right)\!\!,\label{con:transmallercharge}  \\
	&\!\!& ~\eqref{OP3.3P1con:SNR bound}, \eqref{constraint hatP sum}, \eqref{constraint hatP upper bound2}.\nonumber
%	&\!\!& \sum_{j=1}^M \frac{1}{\hat{P_j}}\leq P_{\rm{total}}.\label{con:transmitpower}
	\end{eqnarray}
\end{subequations}

 \vspace{-.5cm}
Afterwards, we propose the following lemma to prove the convexity of   Subproblem (SP1).
  \vspace{-.1cm}
\begin{lemma}\label{le:3.3convexobjprove}
{Subproblem (SP1) is convex}.
\end{lemma}
  \vspace{-.5cm}
%\vspace{-.5cm}
\begin{proof}
We start the proof by verifying the convexity of all constraints. First, constraint \eqref{con:transmallercharge} is conducted via the proposed convex approximation, which is  convex. 
Second,~\eqref{OP3.3P1con:SNR bound} can be represented by two linear constraints, which are  also convex. Third, for  constraint~\eqref{constraint hatP sum} and ~\eqref{constraint hatP upper bound2}, clearly $\frac{1}{\hat{P_j}}$ is convex in $\hat{P_j}$. Noted that $\frac{1}{\hat{P_j}}$ is independent in $\hat{P_{j'}}, \forall {j'} \ne j$ and $\boldsymbol{\bar P}$, thus $\frac{1}{\hat{P_j}}$ is joint convex in $(\boldsymbol{\hat{P}},\boldsymbol{\bar{P}})$. Therefore, $\sum_{j=1}^M \frac{1}{\hat{P_j}}$ is also convex.

Then, the major task to prove Lemma~\ref{le:3.3convexobjprove} becomes showing the convexity of the objective function. 
According to~\cite{SWIPT2019hu}, while $\gamma\geq1$, $\varepsilon_i$ is convex in $\gamma_i$, i.e., $\frac{\partial^2 \varepsilon_i}{\partial \gamma_i^2}\geq 0$. 
Based on~\eqref{3.1 SNR}, both first and second derivatives of $\gamma_i$ are not negative, i.e., $\frac{\partial \gamma_i}{\partial \bar{P}_i} =\frac{ \bar{z}_{\rm2,i}}{\sigma_i^2} \geq 0,\frac{\partial^2 \gamma_i}{\partial \bar{P}_i^2}=0.$
Therefore, the second derivative of~$\varepsilon_i$ with respect to $\bar{P}_i$ is written as 
  \vspace{-.1cm}
\begin{equation}  \vspace{-.1cm}
    \begin{split}
    \frac{\partial^2 \varepsilon_i}{\partial \bar{P}_i^2} & = \frac{\partial^2 \varepsilon_i}{\partial \gamma_i^2} \left(\frac{\partial \gamma_i}{\partial \bar{P}_i}\right)^2\!+\frac{\partial \varepsilon_i}{\partial \gamma_i} 
    \frac{\partial^2 \gamma_i}{\partial \bar{P}_i^2} = \frac{\partial^2 \varepsilon_i}{\partial \gamma_i^2}  (\frac{ \bar{z}_{\rm2,i}}{\sigma_i^2})^2 \,\geq \, 0,
    \end{split}
\end{equation}
% where 
%  According to \cite{ }, while $\gamma\geq1$, the error probability $\varepsilon$ is convex in $\gamma$ , i.e., $\frac{\partial^2 \varepsilon}{\partial \gamma^2}\geq 0$. Moreover, in our problem $\gamma_i=\bar{P}_i \eta_i$, where $\eta_i$ is a positive constant.  We have $\frac{\partial \gamma_i}{\partial \bar{P}_i} =\eta_i \geq 0,\frac{\partial^2 \gamma_i}{\partial \bar{P}_i^2}=0.$
% \begin{equation}
%     \frac{\partial \gamma_i}{\partial \bar{P}_i} =\eta_i \geq 0,\frac{\partial^2 \gamma_i}{\partial \bar{P}_i^2}=0.
% \end{equation}
which indicates that the error probability of single link transmission is convex in the  transmit power of the corresponding sensor, i.e., $\varepsilon_i$ is convex in $\bar{P}_i$.

 In addition,  %$\varepsilon_i$ is not directly affected by $\hat{P}$ 
 %i.e., $ \frac{\partial^2 \varepsilon_i}{\partial \hat{P_i}^2}=0$.
 %Moreover, 
  in the local subproblem, 
$\varepsilon_i  ={\cal P}\left( \gamma_i(\bar{P}_i)\right)$, which is influenced by $\bar{P}_i$ but independent in   $\Bar{P}_{i'}, \forall i' \ne i$, i.e., we have  $ \frac{\partial \varepsilon_i}{\partial \Bar{P}_{i'}} =0$, $ \frac{\partial^2 \varepsilon_i}{\partial \Bar{P}_{i'}^2}=0$, and
   $ \frac{\partial^2 \varepsilon_i}{\partial \bar{P}_i \partial \Bar{P}_{i'}}=0$, case $i'\neq i$.
  Moreover,  $\varepsilon_i$ is not  affected by $\boldsymbol{\hat{P}}$  in the local subproblem.
Hence, it can be shown that the Hessian matrix of~$\varepsilon_i$ to~$(\boldsymbol{\hat{P}},\boldsymbol{\bar{P}})$ has only one non-zero element $\frac{\partial^2 \varepsilon_i}{\partial \bar{P}_i^2}\ge0$, i.e., being semi-positive definite. In other words, $\varepsilon_i$ is jointly convex to $(\boldsymbol{\hat{P}},\boldsymbol{\bar{P}})$. As a sum of convex functions, the   objective function of (SP1) is also jointly convex to $(\boldsymbol{\hat{P}},\boldsymbol{\bar{P}})$, and thus   
Subproblem (SP1) is convex. 
  \vspace{-.1cm}
 \end{proof}

According to the lemma 1,   Subproblem (SP1) can be efficiently solved and the flow of the proposed SCA-based solution is described in Algorithm~\ref{al:algo1}. 
{In Algorithm 1, we first  initialize the values of transmit power $ \boldsymbol {\hat{P}}^{(0)}$. In particular, we utilize an improved initial point (IIP) decision strategy to obtain the initial feasible solutions:
1) Based on the monotonic optimization programming tool, we obtained a set of minimum source power $\boldsymbol P_{\rm{min}}$ that satisfies the constraint~(11b) in Original Problem (OP1), which is defined as basic initial point (BIP) decision strategy. 
2) Calculate the remaining power that has not been allocated to sources (i.e., $P_{\rm{remaining}}=P_{\rm{total}}-\sum_{j=1}^N P_{\rm{min},j}$).
3) The remaining power is equally allocated to all WPT sources, i.e., the initialized transmit power of $\rm S_j$ is given as $P_{{\rm initial},j}=\min\{P_{\rm{min},j}+\frac{P_{\rm{remaining}}}{M},P_{\rm max}\}$.
As a result, $  {\hat{P}_j}^{(0)}=\frac{1}{P_{{\rm initial},j}}$.}
After initialization, the local problem (18) is addressed aiming at minimizing the overall error probability by optimizing $(\boldsymbol {\hat{P}},\boldsymbol {\bar{P}} )$.
In the $\tau$-th iteration, via solving the local problem, we determine the optimal solution $(\boldsymbol {\hat{P}}^{(\tau)},\boldsymbol {\bar{P}}^{(\tau)} )$ of the  problem. Subsequently, in the second step, a new local problem based on $(\boldsymbol {\hat{P}}^{(\tau)},\boldsymbol {\bar{P}}^{(\tau)} )$ arises.
Through the iteration repetition until convergence,  an efficient sub-optimal solution to problem  (P1) will be obtained. %is converged.

{ Finally, we investigate the complexity of the proposed iteration algorithm based on the ellipsoid method. In problem (SP1), the number of optimization variables is $\left(M+N\right)$, resulting in $\mathcal{O}((M+N)^2\frac{1}{\epsilon})$ rounds of ellipsoid updates, where $\epsilon$ denotes the optimization threshold. During each round of ellipsoid update, the complexity cost for object function is $\mathcal{O}((M+N)^2)$ and that for constraints is $\mathcal{O}((M+N)(N+N+M))$. Finally, with $\Phi$ rounds of iteration, the complexity of the proposed algorithm can be obtained as $\mathcal{O}\!\left(\!\Phi(M+N)^2\frac{1}{\epsilon}\!\right)\!\times\Big(\!\mathcal{O}(M+N)^2 \!+\!\mathcal{O}\big((M+N)(N+N+M)\big)\!\Big)=\mathcal{O}\left(\Phi(M+N)^3(M+2N)\frac{1}{\epsilon}\right)$.}
 \vspace{-.1cm}
% \begin{equation}\label{complexity}
%     \begin{aligned}
%     &\mathcal{O}\!\left(\!\Phi(M+N)^2\frac{1}{\epsilon}\!\right)\! \\
%     &\times\Big(\!\mathcal{O}(M+N)^2 
%     \!+\!\mathcal{O}\big((M+N)(N+N+M)\big)\!\Big)\\
%     &\!=\mathcal{O}\left(\Phi(M+N)^3(M+2N)\frac{1}{\epsilon}\right).
%     \end{aligned}
% \end{equation}

\begin{algorithm}[!t]%\small
	\small
	\algsetup{linenosize=\large}
	%\vspace{.05in}
	\caption{\bf{ Power Allocation Algorithm  }}
	\begin{algorithmic}
		\STATE \noindent{\small\bf{$\!\!\!\!\!\!$Initialization}} \\
		
		\STATE   Initialize a feasible $( \boldsymbol {\hat{P}}^{(0)},\boldsymbol {\Bar{P}}^{(0)})$.
		%\STATE $r=0.$
		\STATE   Initialize the overall error probability $ \boldsymbol {\varepsilon_\text{o}}^{(0)}$ based on $(  \boldsymbol {\hat{P}}^{(0)},\boldsymbol {\hat{P}}^{(0)})$.
		%\STATE $r=0.$
	
		 \STATE \noindent{ \bf{$\!\!\!\!\!\!\!\!$Iteration}} \\
		 \STATE \noindent{\bf{a)}}~~ $\varepsilon_{\text{o,min}}=\varepsilon_{\text{o}}^{(0)}$, $\boldsymbol{\hat{P}}_{\rm{opt}}= P^{(0)}$, $\boldsymbol{\bar{P}}_{\rm{opt}}= \bar{P}^{(0)}$, $\tau=0$\\ 
		 
		 \STATE \noindent{\bf{b)}}~~ iteration number $\tau=\tau+1$ \\
		 
	         \STATE \noindent~~ \quad   {\bf{for}} sensor $i=1:N$\\ %${\rm{R}}_i$=1:N\\
	         \quad \quad \quad \quad   {\bf{for}} source $j=1:M$
	     
	         \STATE \noindent{\bf{c)}}\quad \quad \quad \quad \quad \quad Determine parameters $A_{ij}^{(\tau)}$ according to (\ref{A1})  \\
	          \STATE \noindent~~ \quad \quad \quad {\bf{endfor}}  \\
	          
	       \STATE \noindent{\bf{d)}}~~~~~ \quad  Determine parameters $B_i^{(\tau)}$ according to (\ref{B1})\\
	       ~\quad \quad  {\bf{endfor}}
	           
	       \STATE \noindent{\bf{e)}}~~ \quad Update $\boldsymbol{\hat{P_j}}^{(\tau+1)}$ and $\boldsymbol{\bar{P}_i}^{(\tau+1)}$ by solving (18)  \\
   \STATE \noindent{\bf{f)}}~~~~  {\bf{if}} ${\left|\varepsilon_{\text{o}}^{_{({\rm \tau}+1)}}-\varepsilon_{\text{o,min}}\right|} \big/ {\varepsilon_{\text{o,min}}}\leq \varepsilon_{\text{converge-threshold}}$    \\%10^{-8} \\
   \STATE \noindent~~~ \quad\quad \,\,  {\bf{break}} \\
   \STATE \noindent~~~~\quad  {\bf{endif}} 
	   \STATE \noindent{\bf{g)}}~~~~  {\bf{if}} $\varepsilon_{\text{o}}^{({\rm \tau}+1)}<\varepsilon_{\text{o,min}}$ \\
	   \STATE \noindent~~~~  \quad     $\varepsilon_{\text{o,min}}= {\varepsilon_{\text{o}}^{({\rm \tau}+1)}}$, $ \hat{P_{j}}= {\hat{P_j}^{({\rm \tau}+1)}}$,
	
	     \STATE \noindent{\bf{h)}}~~~  {\bf{else}} jump to step b).
	   \STATE \noindent~~ \quad  {\bf{endif}} \\
	\end{algorithmic}
	\label{al:algo1}
\end{algorithm}

 \vspace{-.1cm}
\section{Joint Power and Blocklength  Allocation}
 \vspace{-.1cm}
\label{sec:Power and Blocklength Joint Allocation}
So far, a power allocation design with fixed frame structure has been completed in the previous section. 
In this section, our work is extended to the joint resource allocation design under scenarios with dynamic frame structure, namely the source power  and blocklength allocated for WPT phase and short packets transmission phase are supposed to be jointly optimized %with the objective of 
for minimizing the overall error probability.
% In particular, we first propose an alternating based method.
%s proposed
%, where the original problem is decomposed into two subproblems, which will be alternately solved till the solution converges.
%  Afterwards,  as the second approach for resolving the problem, a direct joint optimization algorithm will be proposed for a high-quality solution with a low computational complexity.

 \vspace{-.3cm}
\subsection{Problem Formulation}
\label{joint op }
In contrast to fixed frame structure, dynamic frame structure permits flexible blocklength adjustment for better system performance achievement based on a variety of influencing factors, including channel states, packet size, transmit power, etc. 
Under such  dynamic frame structure, we aim to minimize the overall error probability under both total power and total blocklength constraints. For this purpose, there exists a tradeoff with respect to the power and blocklength. Power allocation design has been discussed in section~\ref{sec:power allocation}, therefore we  focus on the latter in this subsection. 
One the one hand, there exists a compromise between the blocklength allocated for WPT phase and WIT phase. Allocating more blocklength for WPT phase contributes to higher harvested energy, accordingly higher SNR based on~\eqref{3.1 barP}~\eqref{3.1 SNR}, which helps reduce the overall error probability according to~\eqref{qfunc}, but the blocklength for WIT becomes shorter, which has negative impacts on the overall error probability. Allocating more blocklength for WIT phase can directly improve the FBL reliability according to~\eqref{qfunc}, nevertheless, the harvested energy may become insufficient, i.e., the SNR becomes smaller, which makes negative impacts on system reliability.
One the other hand, there also exists a tradeoff inside the WIT phase, involving the blocklength allocation among multiple sensors. For each sensor, a longer blocklength benefits its own short packet transmission, but  the available  blocklength for other sensors is reduced, which negatively affects other sensors'  reliable transmission. To guarantee the overall system reliability, each short packet must be transmitted reliably, since the overall error probability  to some extent is likely to be determined by the worst transmission.
In addition, although longer blocklength can directly reduce error probability based on~\eqref{qfunc}, with given harvested energy, longer WIT blocklength results in lower sensor transmit power, namely lower SNR according to~\eqref{3.1 barP}~\eqref{3.1 SNR}, which again portrays the tradeoff regarding blocklength. 
Therefore, appropriate blocklength allocation design is of great importance to guarantee the system reliability in a multi-source WPT enabled multiple short packets transmission network.
% In a multi-sensor short packet transmission network with a total blocklength restriction, how to allocate blocklength to multi-sensors is also of great importance for enhancing system  reliability. 
In order to minimize the overall error probability of the system, the joint power and blocklength optimization can be formulated as
\begin{subequations}
	\begin{eqnarray}
		\label{OP1:power_linear}
({\rm OP2}):&\!\!{\min\limits_{\{\boldsymbol{P}\},\{\boldsymbol{m}\},m_0}}\, &\sum_{i=1}^N \varepsilon_i  \label{4.1OPobj:error}\\
    &\rm{s.t.:} &  m_0+\sum_{i=1}^N m_i \leq m_{\rm{total} },\label{4.1OPcon:sum m} \\
    %(\text{constraint missing in (P3)?})
    &\!\!& \frac{{\bar{P}_i}(  Q_{\rm{ch,}i}(\boldsymbol{P}),m_i,m_0) \bar{z}_{\rm2,i}}{\sigma_i^2} \geq  \max\{1, 2^\frac{k_i}{m_i}-1\}, \label{4.1OPcon:SNR}\\
    %\left({\bar{P}_i}(  Q_{\rm{ch,}i}(\boldsymbol{P}),m_i)\to{\bar{P}_i}(  Q_{\rm{ch,}i}(\boldsymbol{P}),m_i,m_0)\right) \\
	&\!\!& \eqref{3.1 barP}, \eqref{OP3.1con:sum P},{\eqref{OP3.1con:upper P}}, \nonumber 
	\end{eqnarray}
\end{subequations}
where constraint~\eqref{4.1OPcon:sum m} announces the upper bound of total blocklength (latency) of the whole network. {To facilitate the problem settlement, we utilize the integer relaxation strategy to treat variables $m_0$ and $\boldsymbol m$ as continuous variables, i.e., $m_0, \boldsymbol m \in \mathbb{R}$}.
In~\eqref{4.1OPcon:SNR}, term ${\bar{P}_i}(  Q_{\rm{ch,}i}(\boldsymbol{P}),m_i,m_0)$ indicates the transmit power of $R_i$, which is jointly affected by $\boldsymbol{P}$, $m_i$ and $m_0$.
% which is whose form is identical to the one in the previous chapter,  i.e.,~\eqref{3.1 barP}. However, evident differences exists in its practical meaning, i.e., besides effects from multiple source power  $\boldsymbol{P}$, ${\bar{P}_i}$ in problem (OP2) is also determined by blocklength $m_i$, i.e., $\bar{P}_i = \frac{Q_{\rm{ch},i}(\bold{P})m_0 }{m_i}$. 
% Accordingly, SNR $\gamma_i$ is also jointly affected by $\boldsymbol{P}$ and $m_i$ based on~\eqref{3.1 SNR}, which implies that~\eqref{OP3.1con:SNR bound} announces restriction on both $\boldsymbol{P}$ and $m_i$ for SNR and Shannon capacity theory limitations.
In~\eqref{3.1 barP}, relationship between $\boldsymbol P$ and $\bar P_i,\forall i=1,\cdots,N$ is depicted.
Constraints~\eqref{OP3.1con:sum P} and~\eqref{OP3.1con:upper P} respectively announce the total power consumption restriction and the upper bound of transmit power for each WPT source.
Recalling~(\ref{qfunc}), i.e., 
 $\varepsilon_i  ={\cal P}\left( \gamma_i, r_i,m_i\right)$, with given packet size $k_i$, the single link transmission error probability $\varepsilon_i$ in~\eqref{4.1OPobj:error} is jointly influenced by $\left(\gamma_i,m_i\right)$.
 Combining~\eqref{Qch_original},~\eqref{3.1 SNR}, the error probability of the transmission from $\rm{R}_j$ to D can be written as
 \begin{equation}
     \varepsilon_i(\boldsymbol{P},m_i,m_0) ={\cal P}\left(\gamma_i\left(\bar{P}_i\left(\boldsymbol{P},m_i,m_0\right)\right),r_i,m_i\right).
     \label{error_joint}
 \end{equation}

Clearly, the nonlinearity in EH process, as well as the complicated Q-function in FBL model, {have} made   Problem (OP2) nonconvex and  intractable. In particular, when the WPT source power levels are supposed to be jointly optimized with the blocklength, the joint convexity of~$\varepsilon_i$ can hardly be expected with respect to $(\boldsymbol P,\boldsymbol m)$. Furthermore, the adjustable blocklengths for both WPT and WIT phases are coupled, i.e., having particularly resulted in significant joint impacts on the transmit power $\bar P_i(\boldsymbol P,m_i,m_0)$ and subsequently on the error probability $\varepsilon_i$, which has introduced extreme analysis difficulty in addressing Problem (OP2).

  \vspace{-.3cm}
\subsection{Efficient Solution to Problem (OP2)}
 \vspace{-.1cm}
In {this} subsection,  we provide a SCA-based approach    efficiently addressing Problem (OP2) via   two steps: %the original nonconvex problem
%In particular, we reformulate it to a tractable convex one via 
First, we decouple the complex relations between the WPT and WIT phases by introducing auxiliary variables, which results in a reformulation to Problem (OP2) ; 
Subsequently, we characterize the reformulated problem, especially the reformulated constraints, and conducting corresponding convex approximations.
%performing variable substitutions to enable applying SCA, e.g., for constraint \eqref{3.1 barP} with equality;
%iii. exploiting SCA while conducting a set of convex approximations to the problem. 
%Specific tasks are shown as follows.
% 妙啊！
 
On the first step, we decouple the complicated relationship between the error probability $\varepsilon_i$  in the WIT phase and the resource allocation decision in the WPT phase $\{\boldsymbol P,m_0\}$, while maintaining such coupled relationship in constraints. 
In particular, we introduce auxiliary power variables $\boldsymbol{\bar{P}}$ to be optimized and establish a new coupling relation {between 
 $\varepsilon_i$ and ${\bar{P}_i}$},
 %with respect to~$\varepsilon_i$ 
 where ${\bar{P}_i}$ is the transmit power of $\rm R_i$ for short packet transmission. Therefore,~\eqref{error_joint} is rewritten as
\begin{equation}
     \varepsilon_i(\bar{P}_i,m_i) ={\cal P}\left(\gamma_i\left(\bar{P}_i\right),r_i,m_i\right),
     \label{error_joint2}
\end{equation}
where error probability $\varepsilon_i$ is completely influenced by $(\bar{P}_i,m_i)$ and is independent from $P_j,\forall j=1,\cdots,M$. This step reduces the number of variables that have a direct affection on $\varepsilon_i$ (from $M+2$ to $2$), thus simplifies the original complex relationship.

To guarantee the consistency of the upcoming reformulated problem and the original problem, the following inequality limitation between new optimization variables $\boldsymbol{\bar{P}}$ and original optimization variables $\boldsymbol{P},\boldsymbol{m},m_0$ must be satisfied, 
\begin{equation}
    \bar{P}_i m_i \leq  F_{\rm{nl}}[\boldsymbol{Q}_i(\boldsymbol{P}) ]m_0, \forall{i=1,\cdots,N},
    \label{P P m constraint}
\end{equation}
which represents that the total consumed energy for short packet transmission during WIT phase should be no larger than the total harvested energy during WPT phase. 
{Considering the monotonic increment property of FBL reliability with respect to sensor transmit power $\Bar{P}$ and WIT blocklength $m$, the optimal reliability can be achieved only when all the resources (including power and blocklength) have been fully utilized and no remaining resource exists, i.e., the equality in~\eqref{P P m constraint} holds, which indicates the equivalence of inequality constraint~\eqref{P P m constraint} and the original equation constraint on the optimal solution to the problem.}

{To facilitate the problem solving, we utilize the same variable substitution in Subproblem (SP1), i.e., $\hat P_i=\frac{1}{P_i},i\in\{1,\cdots,N\}$, which is conducive to the further utilization of SCA.}
As a result, we obtained the reformulated problem (P3) after performing the decoupling step. 
\begin{subequations}
	\begin{eqnarray}
		\label{OP3}
\!\!\!({\rm P3}):&\!\!\!\!\!\!{\min\limits_{\boldsymbol{\hat{P}},\boldsymbol{\Bar{P}},\boldsymbol{m},m_0}}\, &\sum_{i=1}^N \varepsilon_i  \label{2obj:varepsilon}\\
    &\rm{s.t.} &  \bar{P}_i m_i- m_0 F_{\rm{nl}}[\boldsymbol{Q}_i(\boldsymbol{\hat P}) ]\leq 0  ,\forall{i,\cdots,N}, \label{con:transmallercharge2}  \\
	&\!\!& ~\eqref{OP3.3P1con:SNR bound}, \eqref{constraint hatP sum},\eqref{constraint hatP upper bound2}, \eqref{4.1OPcon:sum m} ,\nonumber
%	&\!\!& \sum_{j=1}^M \frac{1}{\hat{P_j}}\leq P_{\rm{total}}.\label{con:transmitpower}
	\end{eqnarray}
\end{subequations}
Clearly,  problem (P3) is not convex as constraint~\eqref{con:transmallercharge2} is non-convex to variables $\boldsymbol{\bar{P}},\boldsymbol{P},\boldsymbol{m},m_0$.

On the second step, we efficiently solve Problem (P3) via applying the SCA technique. 
In particular, we provide   convex approximations to the problem by specifically addressing the nonlinear EH  impact  as well as the  nonconvex term of  products  over variables in  constraint~\eqref{con:transmallercharge2}, respectively.
%To ensure the convexity of the upcoming reformulated problem, constraint~\eqref{P P m constraint} requires to be reformulated into a convex form. However, term $F_{\rm{nl}}[\boldsymbol{Q}_i(\boldsymbol{P}) ]$, which denotes the complicated nonlinear EH process, poses a significant difficulty for such nonconvex-to-convex conversion. 

On the one hand, we focus on the aforementioned difficulty introduced by nonlinear EH process, we utilize the similar method in Section~\ref{sec1.3} to convert $F_{\rm{nl}}[\boldsymbol{Q}_i(\boldsymbol{P}) ]$ into a simple linear form. % with respect to $\boldsymbol{P}$.
Namely, we first perform variable substitution $\hat{P}_j=\frac{1}{P_j},\forall{j=1,\cdots,M}$. 
Then, based on the joint convexity relationship between $F_{\rm{nl}}[\boldsymbol{Q}_i(\boldsymbol{\hat{P}}) ]$ and $\boldsymbol{\hat{P}}$ (discussed in Section~\ref{sec1.3}), the term $F_{\rm{nl}}[\boldsymbol{Q}_i(\boldsymbol{\hat{P}})]$  is reformulated into a linear one, i.e., $\left(\sum _{j=1}^M -A^{(\tau)}_{i,j}\hat{P_j}+B^{(\tau)}_i\!\right)$, where $A^{(\tau)}_{i,j}$ and $B^{(\tau)}_i$ can respectively obtained by~\eqref{A1}~\eqref{B1}, based on which constraint~\eqref{con:transmallercharge2} is reformulated as
\begin{equation}
\bar{P}_i m_i- m_0 \left(\sum _{j=1}^M -A^{(\tau)}_{i,j}\hat{P_j}+B^{(\tau)}_i\!\right) \leq 0  ,\forall{i,\cdots,N}. 
\label{approximation1}
\end{equation}

On the other hand,       constraint~\eqref{approximation1} is still nonconvex.
% we can address the issue in
Then,  the remaining   task is to  further conduct convex approximation to constraint~\eqref{approximation1}, while guaranteeing the convexity of other constraints in Problem (P3). To this end, we introduce  the following variable substitution, i.e., set $\Tilde{P}_i=\sqrt{\bar{P}_i},\Tilde{m}_i=\frac{1}{m_i},\forall{i=1,\cdots,N}$. Thus, non-convex term $ \bar{P}_i m_i$ in      constraint~\eqref{approximation1} is converted into a convex form, i.e., $\frac{\Tilde{P}^2_i}{\Tilde{m}_i}$. 
For $-m_0\!\left(\sum _{j=1}^M -A^{(\tau)}_{i,j}\hat{P_j}+B^{(\tau)}_i\!\right)$, since $m_0$ is guaranteed to be positive and $\left(\sum _{j=1}^M -A^{(\tau)}_{i,j}\hat{P_j}+B^{(\tau)}_i\!\right)$ represents the charged power, which is also positive. Thus, based on the inequality of arithmetic and geometric means, i.e., $ab\leq \frac{1}{2}a^2+\frac{1}{2}b^2$ holds $\forall a,b\in \mathcal{R}^+$ and the equality holds when $a=b$, we can obtain that

\begin{equation}
\begin{aligned}
  -m_0\left(\sum _{j=1}^M -A^{(\tau)}_{i,j}\hat{P_j}+B^{(\tau)}_i\right)& = \frac{m_0}{C_i^{(\tau)}} C_i^{(\tau)} \sum _{j=1}^M A^{(\tau)}_{i,j}\hat{P_j} - B^{(\tau)}_i m_0\\
  & \leq  \frac{1}{2 C_i^{(\tau)}} m_0^2 + \frac{1}{2}C_i^{(\tau)}\left(\sum _{j=1}^M A^{(\tau)}_{i,j}\hat{P_j} \right)^2 - B^{(\tau)}_i m_0,
  \label{SCA2}
\end{aligned}
\end{equation}
where the positive constant $C_i^{(\tau)}$ is given by
\begin{equation}
    C_i^{(\tau)}=\frac{m_0^{(\tau)}}{ \sum _{j=1}^M A^{(\tau)}_{i,j}\hat{P}_j^{(\tau)}},
    \label{C1}
\end{equation}
which guarantees the equality in~\eqref{SCA2} holds at the local point $(\hat{P}_j^{(\tau)}, m_0^{(\tau)})$.
{Obviously, both expression $\frac{1}{2 C_i^{(\tau)}} m_0^2$ and $ - B^{(\tau)}_i m_0$ are convex. Then, we investigate the convexity of $\frac{1}{2}C_i^{(\tau)}\left(\sum _{j=1}^M A^{(\tau)}_{i,j}\hat{P_j} \right)^2.$ 
Expression  $\frac{1}{2}C_i^{(\tau)}\left(\sum _{j=1}^M A^{(\tau)}_{i,j}\hat{P_j} \right)^2$ is convex and monotonically increasing in $\Theta_i$, where $\Theta_i=\sum_{j=1}^MA^{(\tau)}_{i,j}\hat{P_j}$ and $A^{(\tau)}_{i,j}\geq0$ , $\hat{P_j}\geq0$. 
Moreover, expression $\Theta_i=\sum_{j=1}^MA^{(\tau)}_{i,j}\hat{P_j}$ is linear, which is jointly convex in $\hat{\boldsymbol P}$. 
According to the convexity preserving property of composite functions, 
% $f(\boldsymbol{x})=h(g(\boldsymbol{x}))$ is convex if $h(g)$ is convex and increasing in $g$, and $g$ is convex, which indicates that 
expression  $\frac{1}{2}C_i^{(\tau)}\left(\sum_{j=1}^MA^{(\tau)}_{i,j}\hat{P_j} \right)^2$ is jointly convex in $\hat{\boldsymbol P}$. }
As a result, non-convex constraint~\eqref{approximation1} can be reformulated into a convex one.

However,  as a consequence of applying the above variable substitution, original convex constraint~\eqref{OP3.3P1con:SNR bound}   becomes   a non-convex one
\begin{equation}
    \frac{\bar{z}_{2,i}}{\sigma_i^2} \Tilde{P}^2_i \geq  \max\{1, 2^{k_i\hat{m_i}}-1 \},
    \label{SNR constraint non-convex}
\end{equation} 
which further needs to be  reformulated   to a convex form.  
We achieve this by dividing   constraint~\eqref{SNR constraint non-convex} into two  constraints
\begin{equation}
    \Tilde{P}_i\geq \sqrt{\frac{\sigma_i^2}{\bar{z}_{2,i}}},
    \label{4.3.2 conSNRsub1}
\end{equation}
\begin{equation}
    k_i\Tilde{m_i}- \log_2(1+\Tilde{P_i}^2 \frac{\bar{z}_{2,i}}{\sigma_i^2} )\leq 0,
    \label{4.3.2 conSNRsub2}
\end{equation}
where~\eqref{4.3.2 conSNRsub1} is linear and convex with given $\sigma_i^2,\bar{z}_{2,i}$. 
%A sufficient condition making constraint~\eqref{4.3.2 conSNRsub2}   be convex is that $k_i\Tilde{m_i}$ is convex while $\log_2(1+\Tilde{P_i}^2 \frac{\bar{z}_{2,i}}{\sigma_i^2})$ is concave.
Obviously, $k_i\Tilde{m_i}$ is linear and convex. Thus, the follwing task is to prove the  convexity of $-\log_2(1+\Tilde{P_i}^2 \frac{\bar{z}_{2,i}}{\sigma_i^2})$.
%we check  the convexity of $-\log_2(1+\Tilde{P_i}^2 \frac{\bar{z}_{2,i}}{\sigma_i^2})$. 
The second-order derivative of $-\log_2(1+\Tilde{P_i}^2 \frac{\bar{z}_{2,i}}{\sigma_i^2})$ can be given as
\begin{equation}
\label{eq:derivativelog}
\begin{split}
    \frac{\partial^2 [-\log_2(1+\Tilde{P_i}^2 \frac{\bar{z}_{2,i}}{\sigma_i^2})]}{\partial \Tilde{P_i}^2} & = \frac{-2\ln2\frac{\bar{z}_{2,i}}{\sigma_i^2}(1-\Tilde{P_i}^2 \frac{\bar{z}_{2,i}}{\sigma_i^2})}{[(1+\Tilde{P_i}^2 \frac{\bar{z}_{2,i}}{\sigma_i^2})\ln2]^2},
\end{split}
\end{equation}
where term $(1-\Tilde{P_i}^2 \frac{\bar{z}_{2,i}}{\sigma_i^2})$ is non-positive when $\gamma_i\geq1$, i.e., $\Tilde{P_i}^2 \frac{\bar{z}_{2,i}}{\sigma_i^2}\geq 1$ holds, which has been guaranteed in~\eqref{4.3.2 conSNRsub1}. 
Therefore, the second-order derivative of $-\log_2(1+\Tilde{P_i}^2 \frac{\bar{z}_{2,i}}{\sigma_i^2})$ is nonnegative and thus constraint~\eqref{4.3.2 conSNRsub2} is convex.

Based on the above designed convex approximations  and  analysis, i.e.,  from \eqref{approximation1} to \eqref{eq:derivativelog},  
Problem (P3) can be solved via a SCA-based approach. %, Problem (P3) is converted into local sub-problems. 
In particular, in the $\tau$-th iteration of the approach, the local subproblem can formulated as 
\begin{subequations}
	\begin{eqnarray}
({\rm SP3}):&\!\!{\min\limits_{\{\hat{P}_j\},\{\Tilde{P}_i\},\{\Tilde{m}_i\},{m_0} }}\, &\sum_{i=1}^N \varepsilon_i(\Tilde{P}_i,\Tilde{m}_i)   \label{4.3.2OP:objective}\\
    &\rm{s.t.:} & \frac{\Tilde{P}^2_i}{\Tilde{m}_i} + {\frac{1}{2 C_i^{(\tau)}} m_0^2 + \frac{1}{2}C_i^{(\tau)}\left(\sum _{j=1}^M A^{(\tau)}_{i,j}\hat{P_j} \right)^2 - B^{(\tau)}_i m_0} \leq 0, \label{4.3.2con: consume<charge}  \\
    &\!\!& \Tilde{P}_i\geq \sqrt{\frac{\sigma_i^2}{\bar{z}_{2,i}}}, \label{4.3.2con:SNR}\\
    &\!\!& k_i\Tilde{m_i}\leq \log_2(1+\Tilde{P_i}^2 \frac{\bar{z}_{2,i}}{\sigma_i^2} ), \label{4.3.2con:shanoon}\\
% 	&\!\!& \frac{\bar{z}_{2,i}}{\sigma_i^2} \Tilde{P}^2_i \geq  max\{1, 2^{k_i\hat{m_i}}-1 \},\label{4.3.2con:SNR} \\
	&\!\!& \sum_{j=1}^M \frac{1}{\hat{ P_j}}\leq P_{\rm{total}},\label{4.3.2con:sum P} \\
	&\!\!&  {\hat{P_j}} \geq \frac{1}{P_{\rm max}},
     \label{4.3.2con:P upper bound}\\
	&\!\!& m_0+\sum_{i=1}^N \frac{1}{\Tilde{m_i}}\leq m_{\rm{total}}.\label{4.3.2con:sum m}
	\end{eqnarray}
\end{subequations}
%{(1. $m_0$ as a variable is missing below $\min$. 2. $m_0$ is missing in (25e).)}
% In addition, according to~\cite{}, the error probability $\varepsilon_i$ is joint convex in the root of transmit power $\bar{P}_i$ and the reciprocal of blocklength $m_i$, i.e., $\varepsilon_i$ is joint convex in $(\sqrt{\bar{P}_i},\frac{1}{m_i})$. To facilitate characterization, we introduce auxiliary variables $\Tilde{P}_i,\forall{i=1,\cdots,N}$ and $\Tilde{m}_i,\forall{i=1,\cdots,N}$, where $\Tilde{P}_i=\sqrt{\bar{P}_i},\Tilde{m}_i=\frac{1}{m_i}$. After auxiliary variables introduction and SCA application, the original problem is reformulated into several local sub-problems. In the $\tau$-th iteration, the problem is formulated as follows.

% Constraint~\eqref{4.3.2con: consume<charge} indicates that the consumed energy for WIT should be smaller than the harvested energy during WPT. Constraint~\eqref{4.3.2con:SNR} and~\eqref{4.3.2con:shanoon} respectively announces the limitation with respect to SNR and coding rate. Constraint~\eqref{4.3.2con:sum P} and~\eqref{4.3.2con:P upper bound}
% respectively denotes the total power limit for all sources and the upper bound of transmit power from individual sources.
% In~\eqref{4.3.2con:sum m} total blocklength limitation is announced. 
Then, we provide the following key lemma for addressing  Problem (SP3).
 \vspace{-.4cm}
\begin{lemma}
Problem (SP3) is convex. %((25c) is not convex constraint but can be reformed to a convex constraint. Before reformulation, (SP3) cannot be stated as convex. Suggest to direct add the reformulated (25c) to (SP3), or state here ``(SP3) can be optimally solved via convex optimization techniques'' instead of ``(SP3) is convex''.)
\end{lemma}
 \vspace{-.5cm}
\begin{proof}
First, we prove the convexity of the object function. According to~\cite{Joint2022Zhu}, the FBL reliability is joint convex in the root of transmit power and the reciprocal of blocklength, i.e., $\varepsilon_i$ is joint convex in $(\Tilde{P}_i,\Tilde{m}_i)$, where $\Tilde{P}_i=\sqrt{\bar{P}_i}$ and $\Tilde{m}_i=\frac{1}{m_i}$. Moreover, $\varepsilon_i$ is affected by $(\Tilde{P}_i,\Tilde{m}_i)$ but independent in $\Tilde{P}_{i'},\Tilde{m}_{i'},\forall{i'\neq i}$, i.e., we have 
$\frac{\partial^2 \varepsilon_i}{\partial \Tilde{P}_{i'}^2}=0$,
$\frac{\partial^2 \varepsilon_i}{\partial \Tilde{m}_{i'}^2}=0$,
$\frac{\partial^2 \varepsilon_i}{\partial \Tilde{P}_{i}\Tilde{P}_{i'}}=0$,
$\frac{\partial^2 \varepsilon_i}{\partial \Tilde{P}_{i}\Tilde{m}_{i'}}=0$,
$\frac{\partial^2 \varepsilon_i}{\partial \Tilde{m}_{i}\Tilde{P}_{i'}}=0$,
$\frac{\partial^2 \varepsilon_i}{\partial \Tilde{m}_{i}\Tilde{m}_{i'}}=0$.
In addition, $\varepsilon_i$ is not affected by $\boldsymbol{\hat{P}}$ in the local sub-prblem. Therefore, in the Hessian matrix of $\varepsilon_i$ to $(\boldsymbol{\hat{P},\boldsymbol{\Tilde{P}},\boldsymbol{m}},m_0)$, only elements $\frac{\partial^2 \varepsilon_i}{\partial \Tilde{P}_i^2}$, $\frac{\partial^2 \varepsilon_i}{\partial \Tilde{m}_i^2}$,$\frac{\partial^2 \varepsilon_i}{\partial \Tilde{P}_i \partial \Tilde{m}_i}$ are non-zero elements and $\frac{\partial^2 \varepsilon_i}{\partial \Tilde{m}_i^2} \frac{\partial^2 \varepsilon_i}{\partial \Tilde{m}_i^2}-(\frac{\partial^2 \varepsilon_i}{\partial \Tilde{P}_i \partial \Tilde{m}_i})^2\geq 0$ is satisfied according to~\cite{Joint2022Zhu}, i.e., the Hessian matrix is semi-positive definite.
It indicates that $\varepsilon_i$ is joint convex in $(\boldsymbol{\hat{P},\boldsymbol{\Tilde{P}},\boldsymbol{m}},m_0)$. 
As a sum of convex functions, the object function $\sum_{i=1}^N \varepsilon_i$ is also joint convex in $(\boldsymbol{\hat{P},\boldsymbol{\Tilde{P}},\boldsymbol{m}},m_0)$.

Subsequently, we move on to prove the convexity of all constraints. Since term $\frac{\Tilde{P}^2_i}{\Tilde{m}_i}$ is jointly convex in $(\Tilde{P}_i,\Tilde{m}_i)$ and  independent in $(\boldsymbol{\hat{P}},\Tilde{P}_{i'},\Tilde{m}_{i'},m_0),\forall{i'\neq{i}}$, term $\frac{\Tilde{P}^2_i}{\Tilde{m}_i}$ is jointly convex in $(\boldsymbol{\hat{P}},\boldsymbol{\Tilde{P}},\boldsymbol{\Tilde{m}},m_0)$.
Moreover, constraint~\eqref{4.3.2con: consume<charge} is obtained from convex approximation, which is also convex.
% term ${-\frac{1}{2 C_i^{(\tau)}} m_0^2 -\frac{1}{2}C_i^{(\tau)}\left(\sum _{j=1}^M A^{(\tau)}_{i,j}\hat{P_j} \right)^2 + B^{(\tau)}_i m_0}$ is concave,  thus constraint~\eqref{4.3.2con: consume<charge} is convex.
Constraint~\eqref{4.3.2con:SNR} and~\eqref{4.3.2con:shanoon} are the reformulated convex constraints, whose convexity have been proved in the previous discussion.
Constraint~\eqref{4.3.2con:sum P} and~\eqref{4.3.2con:sum m} are the sum of convex functions, which are also convex.
As the result, the object function and all constraints are all convex, thus the convexity of (SP3) is proved.
\end{proof}

%{(Re-describe the algorithm, perhaps briefly.)}

Finally, a joint transmit power and blocklength  design is depicted in Algorithm~\ref{al:joint design}.
In particular, we first initialize the value of $( \boldsymbol {\hat{P}^{(0)}},\boldsymbol{ \Tilde{P}^{(0)}},\boldsymbol{ \Tilde{m}^{(0)}},m_0^{(0)})$, { utilizing the similar strategy for Algorithm 1, i.e., first finding the minimum $\boldsymbol P_{\rm{min}},\boldsymbol m_{\rm{min}},m_{0,\rm{min}}$ satisfying constraint (21C) in Original Problem (OP2) and subsequently equally allocating the remaining power and blocklength resources respectively among sources and among WPT and multi-packet transmissions during WIT.}
Subsequently, in the $\tau$-th iteration, we make out the constant parameters $A_{ij}^{(\tau)},B_{i}^{(\tau)},C_{i}^{(\tau)}$, 
solve   Subproblem (SP3) and obtain the optimal solution $\hat{P}_j^{(\tau)}$, $\Tilde{P}_i^{(\tau)}$, $\Tilde{m}_i^{(\tau)}$, $m_0^{(\tau)}$. 
Then, in the second iteration, a new local problem based on $\hat{P}_j^{(\tau)}$, $\Tilde{P}_i^{(\tau)}$, $\Tilde{m}_i^{(\tau)}$, $m_0^{(\tau)}$ arises.
Through repetition,  an efficient sub-optimal solution to Problem (OP2) is converged. { Based on the ellipsoid method, the computational complexity of Algorithm 2 is given as $\mathcal{O}\!\left(\!\Phi(M+2N+1)^2\frac{1}{\epsilon}\!\right)\!\times\Big(\!\mathcal{O}(M+2N+1)^2 \!+\!\mathcal{O}\big((M+2N+1)(M+3N+2)\big)\!\Big)=\mathcal{O}\left(\Phi(M+2N+1)^4\frac{1}{\epsilon}\right)$. }

\begin{algorithm}[!h]%\small
	\small
	\algsetup{linenosize=\large}
	%\vspace{.05in}
	\caption{\bf{ Joint Power and Blocklength Allocation Algorithm }}
	\begin{algorithmic}
		\STATE \noindent{\small\bf{$\!\!\!\!\!\!$Initialization}} \\
		
		\STATE   Initialize a feasible $( \boldsymbol {\hat{P}}^{(0)},\boldsymbol{ \Tilde{P}}^{(0)},\boldsymbol{ \Tilde{m}}^{(0)},m_0^{(0)})$.
		%\STATE $r=0.$
		\STATE   Initialize the overall error probability $ \boldsymbol {\varepsilon_{\rm{o}}}^{(0)}$ based on $( \boldsymbol {\hat{P}}^{(0)},\boldsymbol{ \Tilde{P}}^{(0)},\boldsymbol{ \Tilde{m}}^{(0)},m_0^{(0)})$.
		%\STATE $r=0.$
	
		 \STATE \noindent{ \bf{$\!\!\!\!\!\!\!\!$Iteration}} 
		 
		 \STATE \noindent{\bf{a)}}~~ $\varepsilon_{\rm{o,min}}=\varepsilon_{\rm{o}}^{(0)}$, $\boldsymbol{\hat{P}_{\rm{opt}}}= \boldsymbol{\hat{P}^{(0)}}$, $\boldsymbol{\Tilde{P}_{opt}}= \boldsymbol{\Tilde{P}^{(0)}}, \boldsymbol{\Tilde{m}_{opt}}= \boldsymbol{\Tilde{m}}^{(0)}, m_{0,opt}=m_0^{(0)}$
		 
		 \STATE \noindent{\bf{b)}}~~
		 {\bf {for}} iteration number $\tau=$1:itern
		 
	         \STATE \noindent~~ \quad \quad  {\bf{for}} sensor $i$=1:N\\
	         \quad \quad \quad \quad   {\bf{for}} source $j$=1:M\\
	     
	         \STATE \noindent{\bf{c)}} \quad \quad \quad \quad \quad  Determine parameters $A_{ij}^{(\tau)}$ according to (\ref{A1})  \\
	          \STATE \noindent~~ \quad \quad \quad {\bf{endfor}}  \\
	          
	       \STATE \noindent{\bf{d)}}~\quad \quad \quad  Determine parameters $B_i^{(\tau)}$ according to (\ref{B1})\\
	       
	       \STATE \noindent{\bf{e)}}~\quad \quad \quad  Determine parameters $C_i^{(\tau)}$ according to (\ref{C1})\\
	       ~\quad \quad \quad  {\bf{endfor}}
	           
	       \STATE \noindent{\bf{f)}}~~ \quad Update $\boldsymbol{\hat{P}}^{(\tau+1)}$, $\boldsymbol{\Tilde{P}}^{(\tau+1)}$, $\boldsymbol{\Tilde{m}}^{(\tau+1)}$, $m_0^{(\tau+1)}$ by solving (SP3)  \\
	       \STATE \noindent{\bf{g)}}~~ \quad {\bf{if}} ${\left|\varepsilon_{\text{o}}^{_{({\rm \tau}+1)}}-\varepsilon_{\text{o,min}}\right|} \big/ {\varepsilon_{\text{o,min}}}\leq \varepsilon_{\text{converge-threshold}}$    \\%10^{-8} \\
   \STATE \noindent~~~ \quad\quad \,\,  {\bf{break}} \\
   \STATE \noindent~~~~~\quad  {\bf{endif}} \\
   
	   \STATE \noindent{\bf{h)}}~~~~~ {\bf{if}} $\varepsilon_o^{({\rm itern}+1)}<\varepsilon_{o,min}$ \\
	   \STATE \noindent{\bf{i)}}~~~ \quad \quad   {\bf{then}} $\varepsilon_{o,min}= {\varepsilon_o^{({\rm itern}+1)}}$, $\boldsymbol{\hat{P}_{opt}}= \boldsymbol{\hat{P}}^{({\rm itern}+1)}$, \\\quad \quad \quad \quad \quad \quad \quad $\boldsymbol{\Tilde{P}_{opt}}= \boldsymbol{\Tilde{P}}^{({\rm itern}+1)}, \boldsymbol{\Tilde{m}_{opt}}= \boldsymbol{\Tilde{m}}^{({\rm itern}+1)}$, $m_{0,opt}= m_0^{({\rm itern}+1)}$
	   \STATE \noindent~~ \quad \quad {\bf{endif}} \\
	   \STATE \noindent~~ \quad {\bf{endfor}} \\
		
	\end{algorithmic}
	\label{al:joint design}
\end{algorithm}
\vspace{-.3cm}
\section{Numerical Results}
\vspace{-.1cm}
 \label{sec:numerical results}

%{(Simulation highlights)}
% 想尝试figure2-5分为一部分，figure6-9分为一部分，但感觉后半部分目的很杂/广，没想好怎么凝练标题
% 可以试试
In this section, we validate our proposed resource allocation designs and evaluate the system reliability via simulations.
Unless specifically mentioned otherwise, the parameter setups of all the simulations are given as follows:
D is located at (0,0) in the Cartesian coordinate. Three WPT sources are deployed on an arc $20$ meters from the D. Five sensors are evenly distributed on a circle with radius $2$ meters and  the center at $(-15,-15)$.
Blocklength of WPT phase and WIT phase are respectively set to $m_0=500$ and $m_{\rm{total}}=2000$. {The duration of a single symbol is set to $T_s=66.7\mu s$.} The upper
bound of the transmit power for each WPT source is set to $30dBm$. The sizes of short packets are set to $90$ bits and the noise power is set to {$-100\rm dBm$. }
{About channel gains, we consider the Rayleigh distribution with the scale factor $\sigma=1$. }
{With respect to the EH process, the parameter setups are adopted based on the circuit components from~\cite{Waveform2016Clerckx}: }the reverse bias saturation current at diode is set to $I_{\rm s}=5\mu A$, the load resistance is set to $\hat{R_{\rm L}}=200\Omega$, and the truncation order is set to $n_0=4$.  {The diode ideality factor $n=1.05$ and thermal voltage $v_t=25.86mV$.
Constants $a=7.37 \times 10^{-3}\rm A^{-1}$, $\beta_2=0.6782, \beta_4=1.53\times10^4$.
Waveform factors are set as $\lambda_{j,2}=1, \lambda_{j,4}=1.5,\forall j\in\{1,\cdots,M\}$. }
{ In addition, it is worth mentioning that the blocklength optimization results (fractional solutions) have been converted to the integers (closest to the fractional solutions). Considering that the practical short blocklength for short packet communication is usually larger than 100 and even much larger, converting our results to integers will only have very minor effects on system reliability performance.}

\vspace{-0.3cm}
\subsection{Validation}
\begin{figure}[!t]
\centering
\includegraphics[width= 0.58\textwidth, trim=2 30 10 70]{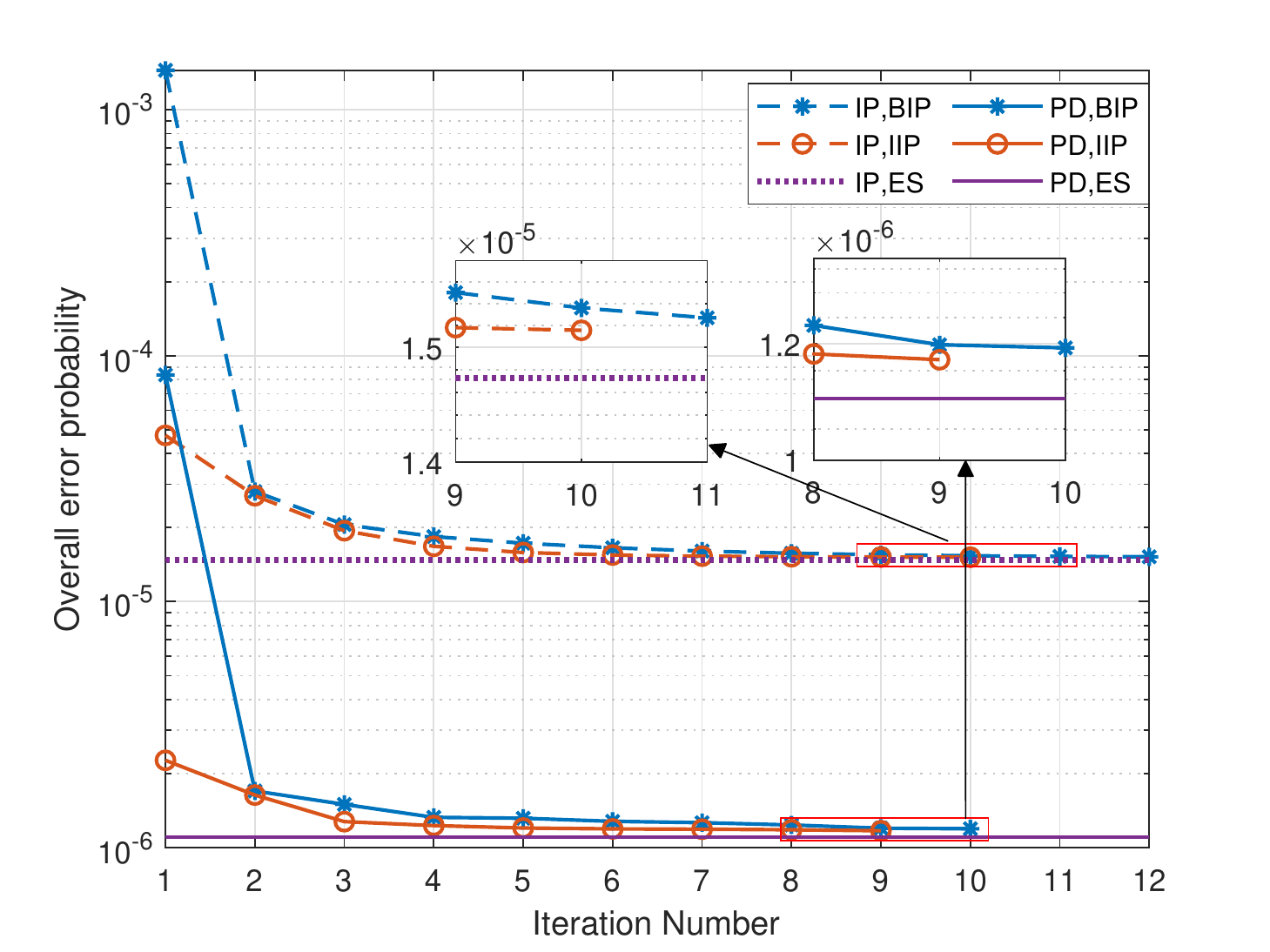}
\DeclareGraphicsExtensions.
\caption{Optimality and convergence performance validation.
%The results of both proposed design (PD) and exhaustive search (ES) are provided.
}
\label{alternating vs joint vs exhuast}
\vspace{-0.8cm}
\end{figure}

{We start with Fig.~\ref{alternating vs joint vs exhuast} to validate the optimality and the convergence performance of our proposed power allocation (IP) and joint resource allocation algorithm (PD). The impact of initial feasible points on optimality is also investigated.
Optimal results obtained from exhaustive search (ES) are provided for comparison. }
Considering the high complexity of the exhaustive search algorithm, only 3 sensors are set for short packet transmission.
{In Fig.~\ref{alternating vs joint vs exhuast}, 
%'IP', `PD' and `ES' respectively represent our proposed power allocation design, joint resource allocation design and the exhaustive search design. 
`BIP' and `IIP' respectively represent the basic initial point decision regime and  the improved initial points decision regime (introduced in Section III and Section IV). 
% Details about initial feasible solutions regimes have been provided in Section III and Section IV. 
Obviously, the overall error probability of iterative algorithms gradually declines as the iteration progresses and the result converges after around 10 rounds of iteration, which indicates the high efficiency of our algorithms. 
Moreover, the results of our proposed iterative algorithm converge to a point, which is very close to the global optimal solution obtained by ES algorithm. In particular, the overall error probability gap between `IIP' and `ES' are  4.2710e-07 and 6.9300e-09 respectively for power allocation design (IP) and joint power and blocklength allocation (PD), which indicates the optimality of our design. The results also verify the equivalence of the optimal solutions to the original problem and the slacked one. 
Moreover, by comparing `BIP' and `IIP', we can learn that the suboptimal solutions obtained by the two regimes (with different initial feasible points) are very close, but `IIP' converges with relatively fewer iteration number.
This result indicates that different initial points will lead to different converge speed but only slight differences in the corresponding suboptimal solutions, which implies a high adaptability of our proposed algorithm with respect to initializations.
%the corresponding suboptimal solutions are close to each other.
}
% Compared with curves of ES, we learn that the efficient sub-optimal solution of PD is close to the optimal one, which confirms the validity of the iterative algorithms.
% Moreover, it is also observed that larger total source power or longer total blocklength helps the network achieve a better reliability since more energy
% is harvested during WPT phase, which enables more power for short packet transmissions in the WIT phase and longer WIT blocklength can directly improve system reliability based on~\eqref{qfunc}.

\begin{figure}[!t]
\centering
\includegraphics[width= 0.58\textwidth, trim=2 25 10 20]{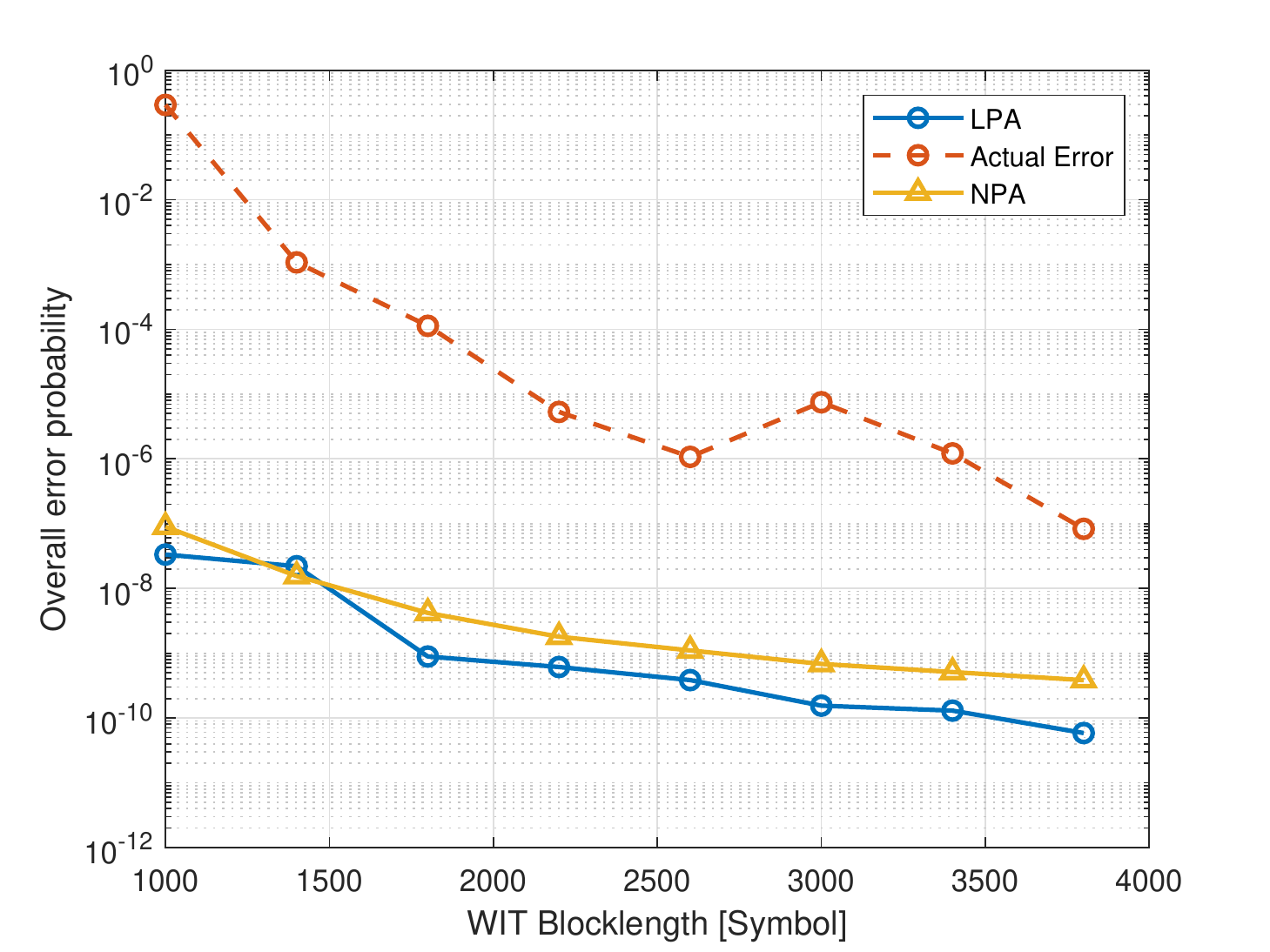}
\DeclareGraphicsExtensions.
\caption{Comparison of system performance based on the linear and nonlinear EH models.} %(typo 'actual' in legend and also in text?)}
\label{linear vs nonlinear (pure power allocation)}
\vspace{-0.9cm}
\end{figure}

Subsequently, we intend to validate the necessity of utilizing the  nonlinear EH model to portray the EH process.
%study how diverse EH models affect the system design and accordingly influence the system reliability.
% In order to focus on the effect of different EH models on the overall error probability, 
Fig.~\ref{linear vs nonlinear (pure power allocation)} depicts the power allocation results based {on both linear and nonlinear EH model} with given WPT blocklength and WIT blocklength, thus the impact of blocklength on system performance  can be circumvented.
In Fig.~\ref{linear vs nonlinear (pure power allocation)}, `LPA' and `NPA' respectively indicate the linear EH model based power allocation and the nonlinear EH model based power allocation. `Actual Error' refers to the practical overall error probability of LPA, i,e., the specific values of multi-source power (obtained from LPA) are brought into the nonlinear EH model to calculate the corresponding practical error probabilities based on~\eqref{qfunc}.
As shown in the figure, the reliability gap between `LPA' and `NPA' is less than half order of magnitude.
However, `LPA' is too idealistic, whose overall error probability is over two orders of magnitude lower than the actual result. 
This can be explained by the fact that the energy conversion efficiency in the actual EH model initially increases and subsequently declines as the received power increases due to the turn-on voltage and the reverse breakdown of the diode(s) utilized in the rectifier circuit.
Nevertheless, linear model fails to accurately model the above nonlinear behavior due to the ideal and simplistic assumption of the constant conversion efficiency.
Therefore, the resource allocation solutions based on the linear EH model have a significant likelihood of not being practically optimal, especially in multi-source wireless powered communication networks, where the received RF signals at each sensor may fluctuate significantly to each other due to diverse channel gains.
%there is a high probability that the resource allocation solutions based on the linear EH model are not practically optimal.
More interestingly, when WIT blocklength increases from 1000 bits to 1800 bits, curve LPA first decreases slowly then the rate of decline fastens.
However, the situation is reversed for the `actual error'. 
The reason could be that under the linear EH model,  more power is allocated to the `most suitable sources' compared with the nonlinear one. %, i.e., a relatively more extreme power allocation scheme is obtained under the linear EH model. 
Nevertheless, while incorporating the above resource allocation solutions into the nonlinear EH model, practical reliability enhancement might be much lower compared with the ideal linear one and even a negative influence may be brought to system performance (inverse increase of error at 3000 blocklength in `Actual Error') since energy efficiency is rather lower with relatively higher received power. 
As a result, with same resource allocation setups, linear and nonlinear models may lead to rather different or even opposite performance trends.
Beyond that, it can also be observed that the `NPA' curve first decreases then flattens as the WIT blocklength increases.
On the one hand, larger  WIT blocklength directly contributes to lower the error probability. 
On the other hand, with certain energy harvested in the WPT phase,  a longer WIT blocklength actually leads to a lower WIT transmit power and accordingly results in a lower SNR, which further negatively contribute{s} to the reliability. 
In other words, {the results also provide an insight into the tradeoff introduced by  WIT blocklength}, which indicates that increasing the WIT blocklength is not always an efficient way to improve the reliability. 

\vspace{-0.1cm}
\begin{figure}[!t]
\centering
\includegraphics[width= 0.58\textwidth, trim=2 25 10 40]{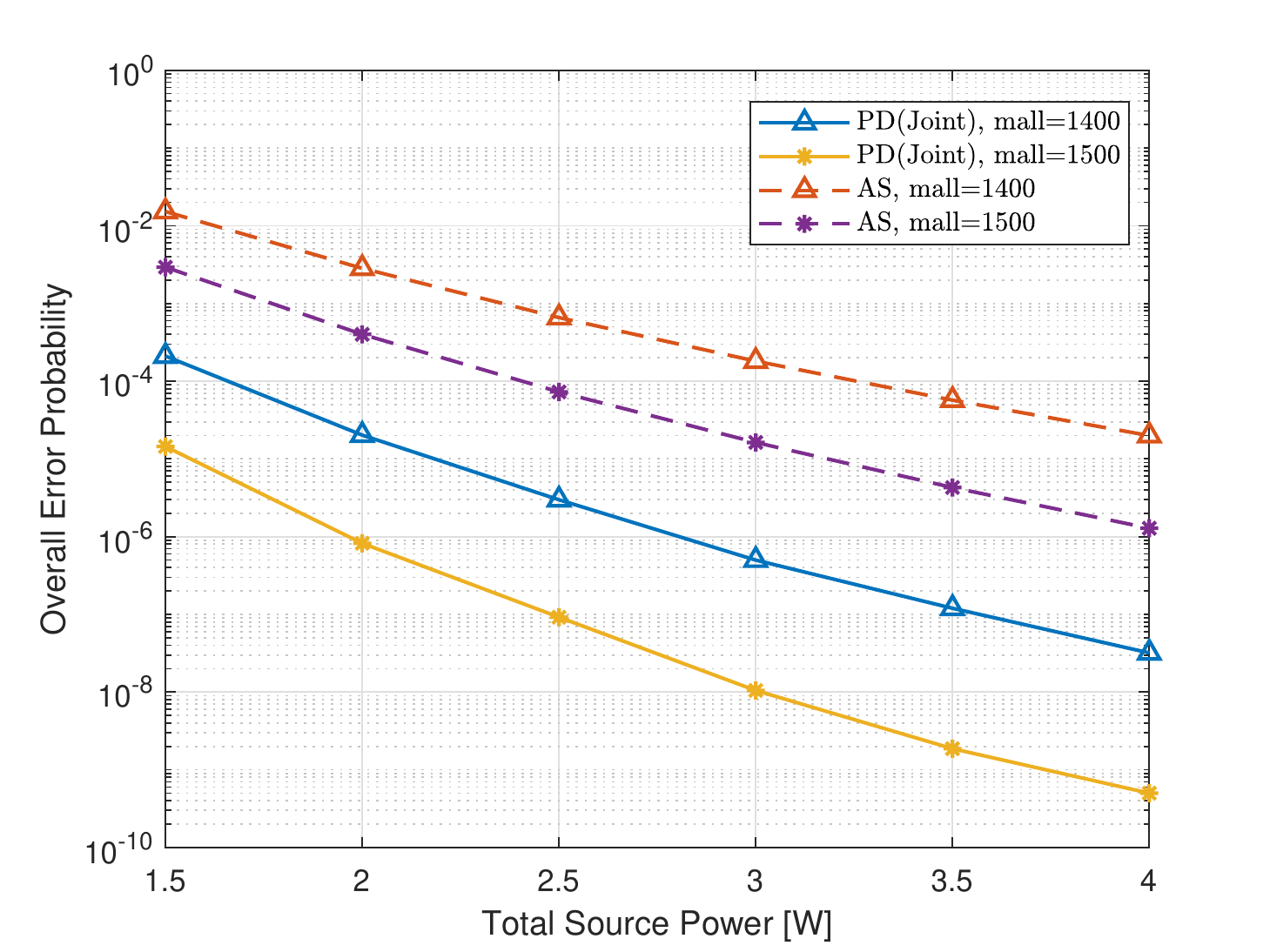}
\DeclareGraphicsExtensions.
\caption{Overall error probability over total source power.}
\label{4-nln-power}
\vspace{-0.9cm}
\end{figure}
\vspace{-0.4cm}
\subsection{Performance Evaluation }

In this subsection, we analyze the system performance under diverse setups with respect to total power (constraint), total blocklength (constraint), packet size and so on.
First, we study the influence of total source power on the system reliability.
As shown in Fig.~\ref{4-nln-power}, higher total transmit power contributes to higher system FBL reliability, which is consistent with the discussion about Fig.~\ref{alternating vs joint vs exhuast}.
Moreover, all curves decrease steadily with virtually constant slopes. It indicates that although EH is a nonlinear process, the  total amount of multiple transmit power  has an approximately loglinear effect on the system reliability.
% All lines in the image have a somewhat constant gradient, which indicates that total transmit power has an approximately linear effect on system stability.
To evaluate the performance enhancement of our proposed design, results based on a resource average sharing strategy, abbreviated as `AS', are also provided.
% strategy a resource average sharing strategy is also provided, which is dented by 'Equal'. 
Through performance comparison, our proposed joint power and blocklength allocation design outperforms the AS and the reliability improvement progressively increases while the total source power increases from $31 dBm$ to $36 dBm$. 
In addition, when total blocklength increases from 1400 to 1500, both AS and PD achieve lower error probability and more interestingly, our PD can enlarge this system reliability enhancement, which indicates that the more resource is available in the system, the more flexible our PD is and the more sufficient resource utilization is achieved, thus the more potential is provided  for reliability enhancement.
%the necessity of PD, especially under scenarios with long blocklength transmission.
%relatively undemanding latency (blocklength) constraints.
% the magnitude of system reliability enhancement  
% % contributed by our design compared with the 'Equal' one 
% is almost unchanged while the total transmit power varies from $32$dBm to $35.5$dBm, which indicates that our proposed nonlinear EH model based joint resource allocation design can provide stable performance enhancement regardless of various total source power levels. 
% In addition, performances based on diverse WIT blocklength setups are also displayed.
% we can learn that larger WIT blocklength helps reduce the overall error probability, however, the degree to which it improves the system reliability is decreased compared with the shorter WIT blocklength.

 \begin{figure}[!t]
\centering
\includegraphics[width= 0.58\textwidth, trim=2 25 10 50]{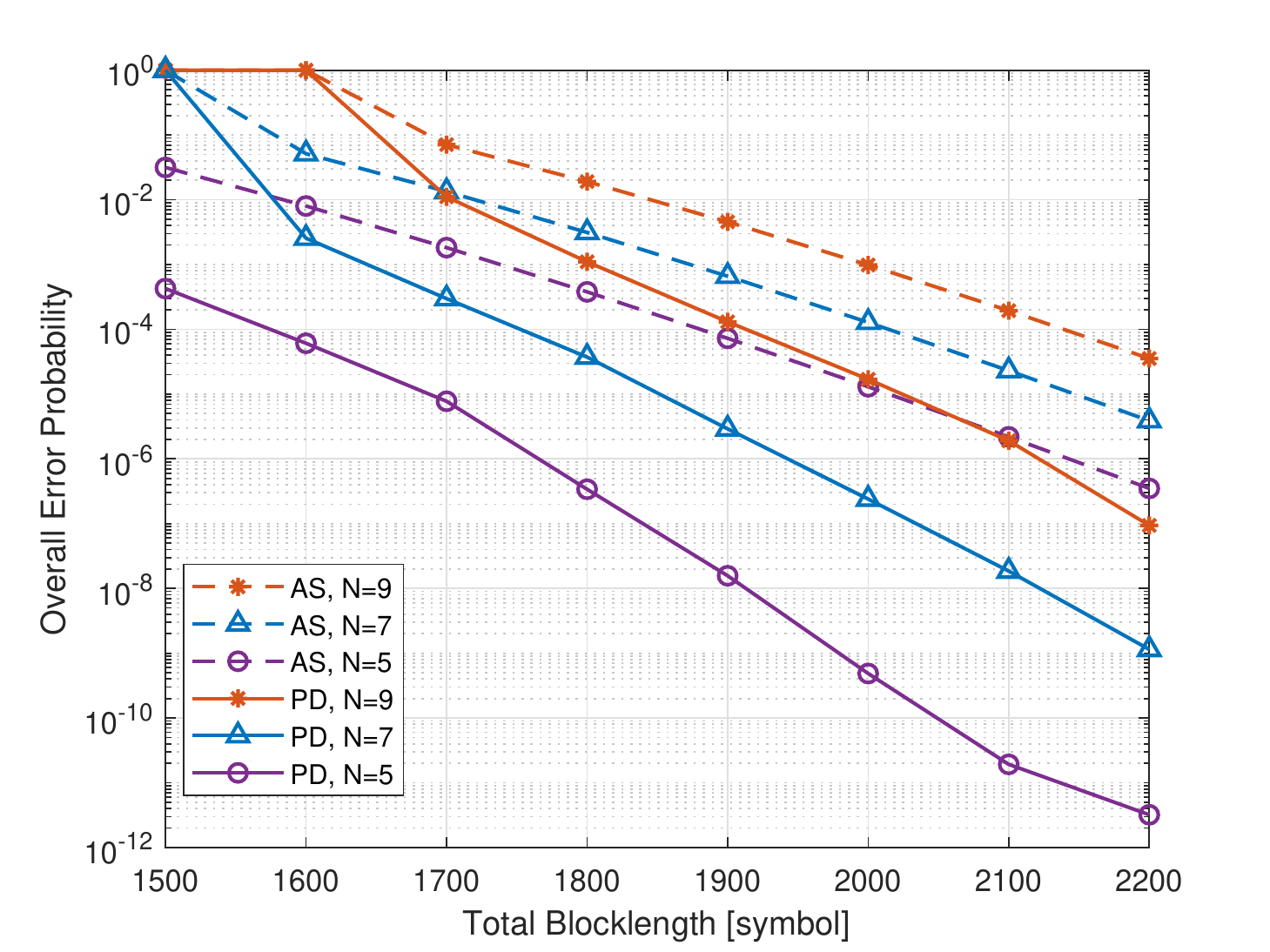}
\DeclareGraphicsExtensions.
\caption{Overall error probability over WIT blocklength with different sensor number.}
\label{4-nln-blocklength}
\vspace{-0.9cm}
\end{figure}
{For the sake of a clearer description of the impact of blocklength on the system reliability, in Fig.~\ref{4-nln-blocklength} we depict the relationship between error and WIT blocklength under various sensor number.}
Clearly, the overall error probability monotonically decreases in the WIT blocklength, i.e., longer blocklength contributes to higher reliability. 
Moreover,  in all cases (various sensor number and total blocklength), our PD outperforms the AS in terms of system reliability in an evident manner.
More interestingly, this performance enhancement varies under diverse cases. For example, when the total blocklength ranges from 1500 to 2200, the performance enhancement boosts from two orders of magnitude to almost six orders   while five sensors are deployed, i.e., longer blocklength contributes to more significant reliability enhancement. 
When the blocklength is 2100, the performance enhancement decreases from three orders of magnitude to two while sensor number enlarges from 5 to 9, which indicates that fewer sensors lead to relatively more significant reliability enhancement.
The aforementioned phenomena can be explained by the fact that {having fewer sensors or setting a longer blocklength likely  provides  additional flexibility in system design, resulting in a greater potential of the performance enhancement  through the resource allocation.} 
Specifically, the overall error probability is approximated as the cumulative sum of error probabilities of all short packet transmissions, i.e., a extremely poor transmission is highly likely to have a decisive impact on overall error probability bottleneck, resulting in  very low system performance. 
To avoid such extreme situation, each sensor requires to be allocated sufficient blocklength for short packet transmission.
Therefore larger total blockelngth or fewer sensors provides more room for system performance enhancement since blocklength with higher upper limit is available for
% a larger range of blocklength is available for allocation
per sensor, i.e., more room for blocklength optimization.
In addition, curve `PD, N=9' and curve `AS, N=5' intersect at a point around blocklength 2100, which indicates that our PD can support more sensors' short packet transmission compared with the AS in the same scenario with same system reliability requirements.

%and more interestingly, the curves first decrease rapidly and then flatten as the WIT blocklength increases. 

\begin{figure}[!t]
\centering
\includegraphics[width= 0.58\textwidth, trim=2 20 10 50]{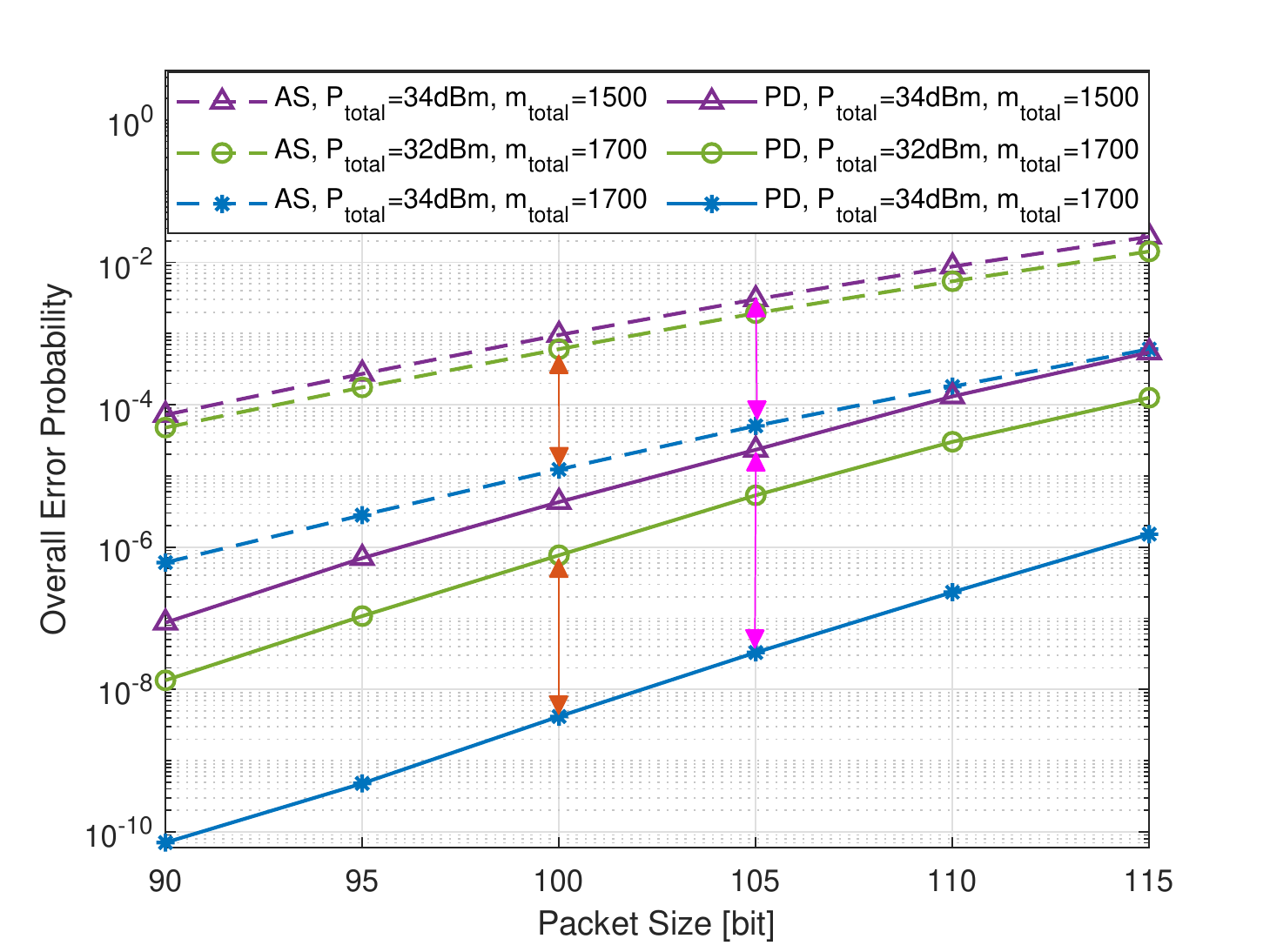}
\DeclareGraphicsExtensions.
\caption{Overall error probability over packet size under nonlinear EH model.}
\label{packetsize}
\vspace{-0.8cm}
\end{figure}

Subsequently, we move on to study the relationship between the overall error probability and the packet size in Fig.~\ref{packetsize}. 
For the sake of analysis, we assume sensors transmit packets with same data amount. 
As the packet size increases, the overall error probability rises monotonically and the slope of each curve rarely changes.
% indicating that transmitting larger packets will accordingly diminish transmission reliability, and the negative impact of the increasing package size on reliability is nearly linear.
% It suggests that transmitting larger packets will diminish system reliability. 
This indicates    that the system can support large packet transmission but  at the price of influencing the   reliability, while  
%system support larger packet transmission,
%transmission of larger packets is possible, but at the expense of system reliability and 
such influence % of the increasing packet size 
on reliability shows an approximate loglinear behavior  in the packet size.
In comparison to the AS benchmark, 
applying the  PD introduces significant  reliability improvements for the system under all circumstances. More interestingly,  such error probability reductions are not constant. In particular,  it is more significant in the small packet size region.   
%but this performance enhancement gradually shrinks as the packet size increases.
Moreover, larger total source power or longer total blocklength contributes to lower error probability and such reliability enhancement introduced by PD is greater than the one by AS, which indicates that when additional resources are available in the system, our PD introduces more flexibility, i,e., a fuller and more efficient utilization of the resources is performed, thus magnifying the reliability enhancement with a given level of resource augmentation.
% the packet size has an almost linear effect on the FBL reliability.
% Further more, besides WIT phase, increasing the blocklength of the WPT phase can also improve the FBL reliability since larger harvested energy are obtained with longer WPT phase.
% In particular, as the WIT blockelngth increases 20\%, i.e.,from 2500 to 3000, the reliability is improved around only one order of magnitude. However, the performance improvement is much more significant (around 2 orders of magnitude) by enhancing 10\% of the WPT phase, i.e., from 500 to 550.
% The result indicates that allocating relatively more time/blocklength for the WPT phase for frame designs is beneficial to system reliability.

\begin{figure}[!t]
\centering
\includegraphics[width= 0.58\textwidth, trim=2 25 10 0]{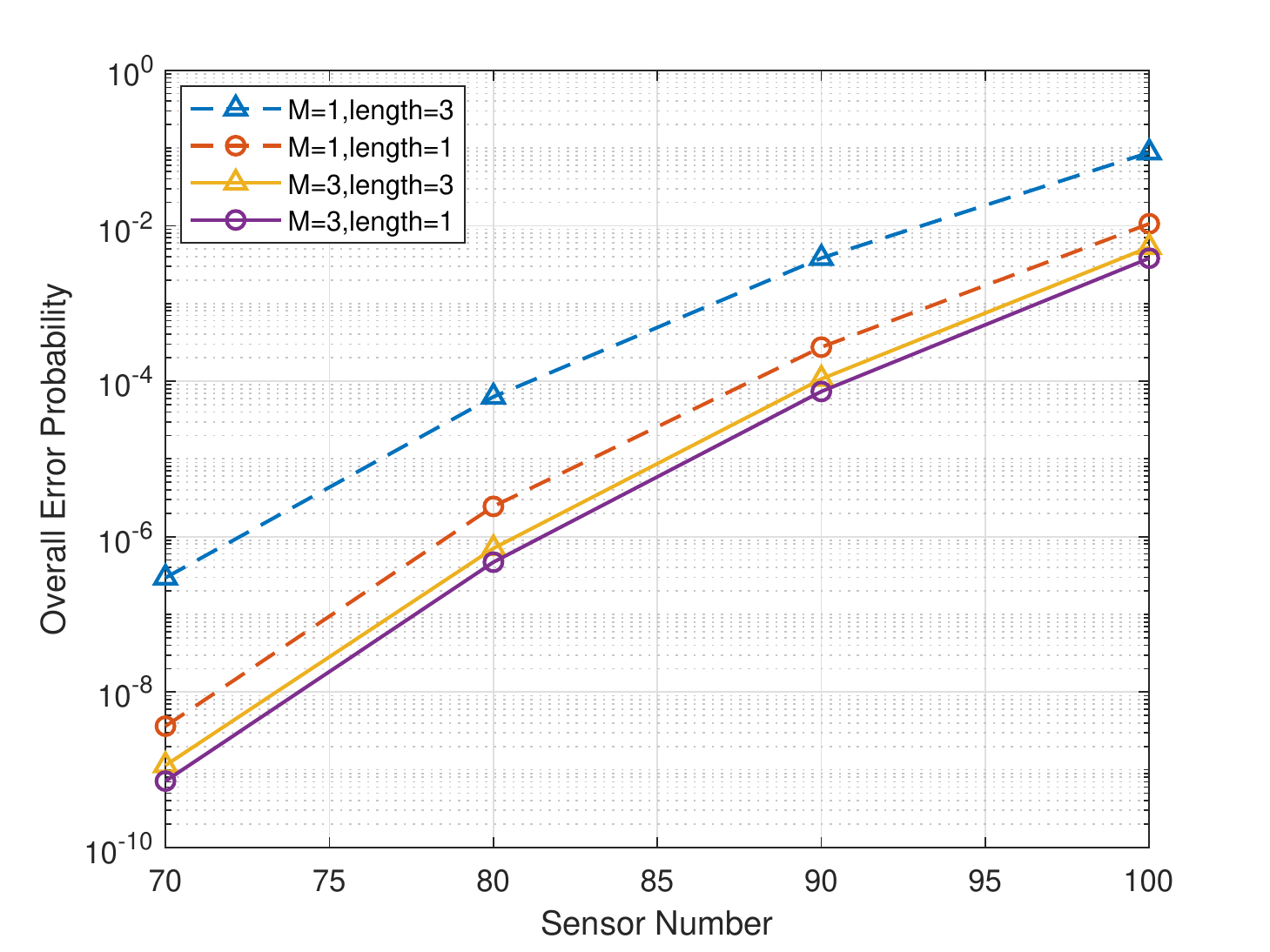}
\DeclareGraphicsExtensions.
\caption{The impacts of the network topology on the overall error probability. }
\label{topology}\vspace{-0.9cm}
\end{figure}
{Our design is of significant generality that can be well adapted to networks with different topologies (including the distribution and number of nodes), based on which we explore the scalibility of our proposed algorithm and the impacts of topology on system reliability.
% Next, we analyze the impact of topology on the system reliability. 
 Fig.~\ref{topology} depicts the reliability performance when more sensors (70 to 100) are deployed for short packet transmissions, which indicates that our algorithm is of significant scalibility that can support massive number of sensors' reliable transmission. As the sensor number increases, the FBL reliability decreases. On the one hand, the total error probability is the sum of transmission error probability of all sensors, which obviously increases with the number of sensors. On the other hand, considering the limited blocklength resource, more sensors represent fewer blocklength assigned per sensor. 
 % the blocklength resource is limited, i.e., more sensors will result in fewer blocklength assigned per sensor. 
 Moreover, as the number of sensors increases, the convergence speed (consumed time for optimization) will become relatively slower to some extent. The result indicates that the network has limited capability to support more sensors' reliable short packet transmissions. 
Subsequently, we compare the performances under single-source and multi-source WPT networks and investigate the impacts of distributions of sensors on system reliability.}
In particular, we randomly deploy the sensors in square with certain length, and change the length of square.
% As shown in Fig.~\ref{topology}, an increasing number of sensors leads to higher FBL error probability, which indicates that the network has limited capability to support more senors' reliable short packet transmissions.
{In comparison to the single-source case,  multi-source WPT significantly improves the  reliability even with the same total amount of WPT energy and WIT blocklength for consumption. 
More interestingly, for single-source case, when sensors are more dispersively distributed, i.e., `length' increases from $1m$ to $3m$, less energy is harvested by certain sensors due to their further distances away from the source, which leads to a decrease in system reliability. 
Nevertheless, such reliability loss is extremely small in the multi-source case, which confirms a great advantage of multi-source WPT, i.e., the  multi-source  diversity   introduces robustness to the reliability performance  with respect to senors locations. }
%similar performances are obtained with diverse  scales of sensor deployment area, which is quite different from that of the single-source case. 
%This results confirm a great advantage of multi-source WPT, i.e., the  multi-source  diversity   introduces robustness to the reliability performance  with respect to senors locations. 

\begin{figure}[!t]
\centering
\includegraphics[width= 0.58\textwidth, trim=2 15 10 40]{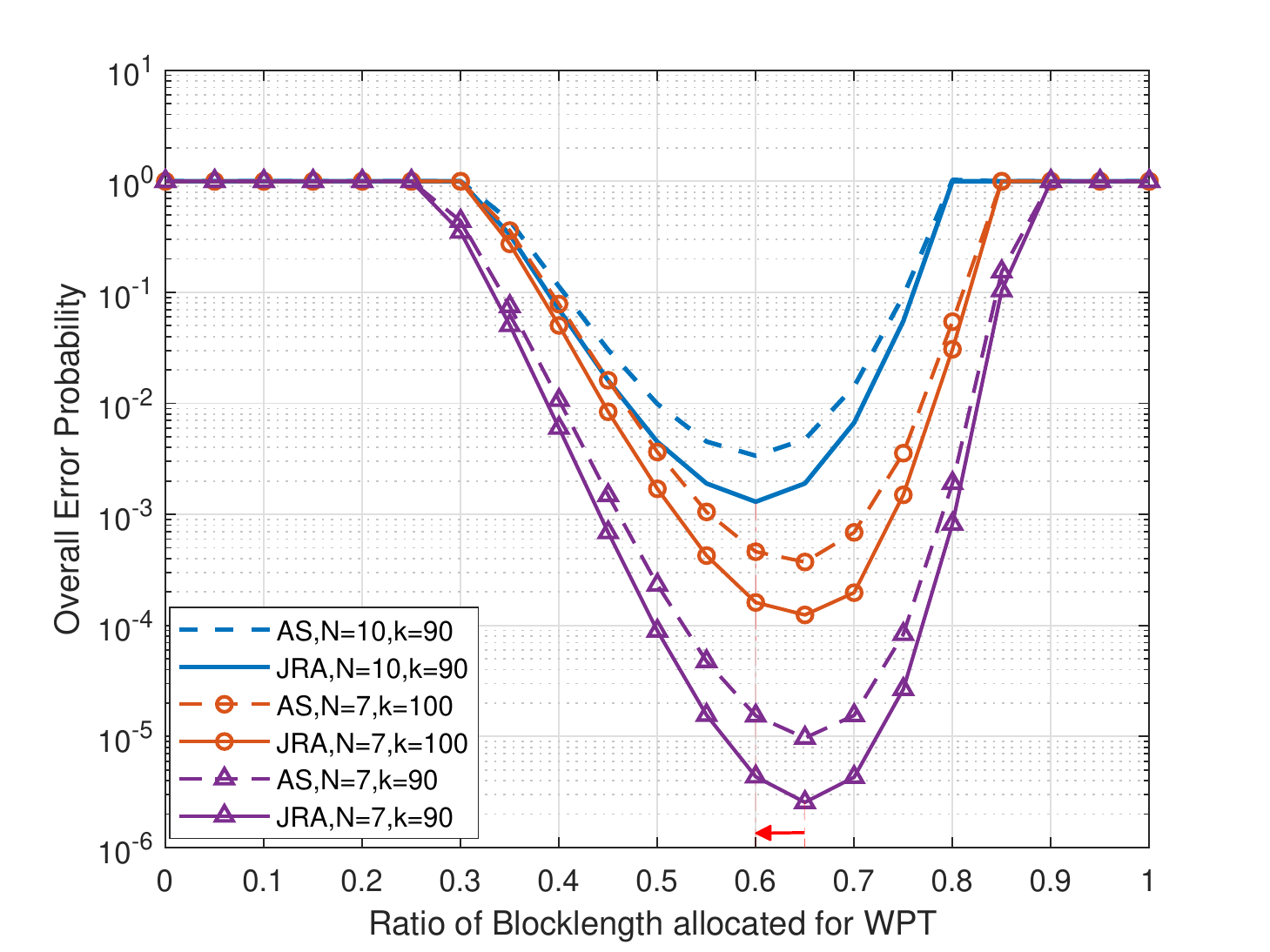}
\DeclareGraphicsExtensions.
\caption{System reliability under diverse normalised WPT blocklength. }
\label{WPT vs WIT blocklength}
\vspace{-0.9cm}
\end{figure}

\vspace{-0.4cm}
\subsection{Performance Advantages of the Joint Design}
\vspace{-0.1cm}
In this subsection, we illustrate the specific performance advantages of  joint power and blocklength allocation design.
Recall that as  discussed in Section~\ref{sec:Power and Blocklength Joint Allocation}, there exists a tradeoff with respect to blocklength in the design, i.e., the compromise between the blocklength allocated for WPT and WIT phase along with the tradeoff inside the WIT phase (blocklength allocation among multi-sensors).
% In Fig.~\ref{4-nln-blocklength}, the impact of the total blocklength on overall error probability has been depicted, while reasonable blocklength allocation solutions for high reliability achievement cannot be directly observed.
To explore the blocklength allocation characteristics, Fig.~\ref{WPT vs WIT blocklength} clearly depicts the system reliability versus diverse normalised WPT blocklength with the given total blocklength, where `JRA' indicates the joint resource allocation. %with given normalised WPT blocklength.
From Fig.~\ref{WPT vs WIT blocklength}, we   learn that the overall error probability is quasi-convex with respect to the normalised WPT blocklength. 
In particular, the error probability first declines and gradually rises when   the normalised WPT blocklength  increases.
% is degraded when the normalised WPT blocklength increases to a certain level, and then it gradually rises when   the normalised WPT blocklength exceed a certain level. 
When the normalised value varies by just 0.2 (from 0.65 to 0.85), the system reliability is weakened by more than four orders of magnitude (from $10^{-5}$ to $10^{-1}$), which indicates that reasonable blocklength allocation between two phases is significant for reliable transmission achievement.
Moreover, for this given system ($N=7,k=90$), the optimal solution of blocklength allocated for WPT phase is higher than that for WIT phase, i.e., 65\% of the total blocklength for WPT and 35\% for WIT, which suggests that relatively more blocklength (time) should be allocated for EH, thus guaranteeing sufficient energy for short packet transmissions.
More interestingly, the optimal value of normalised WPT blocklength decreases from 0.65 to 0.6 when more sensors are utilized (from 7 to 10). This may be explained by the fact that more sensors lead to a higher demand of blocklength for short packet transmisson, since each sensor has a minimum blocklength requirement to guarantee its reliable transmission based on~\eqref{qfunc}, so that a high overall system reliability can be guaranteed.
%that the bottleneck of the system reliability can be guaranteed to be at a high level.
In addition, JRA obviously outperforms AS, and the overall error probability reduction generated by JRA becomes more significant as the packet size decreases (consistent with the results in Fig.~\ref{packetsize}) and the number of sensors increases. The results indicate that deploying more sensors actually introduces a higher diversity, 
%indicates that more variables can be optimized,
i.e., providing more potentials for   system performance enhancements.

\begin{figure}[!t]
\centering
\includegraphics[width= 0.58\textwidth, trim=2 25 10 45]{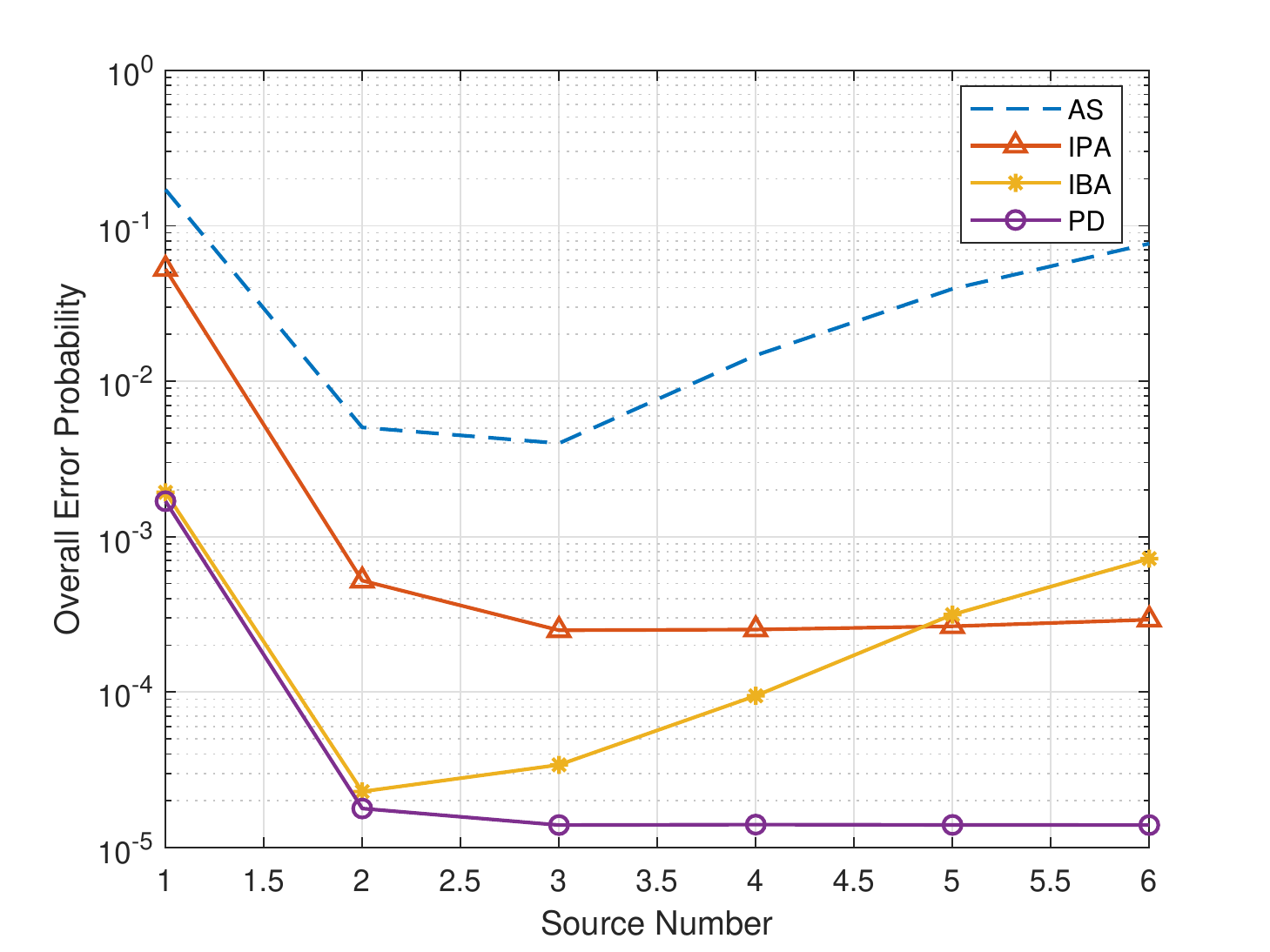}
\DeclareGraphicsExtensions.
\caption{Joint optimization design vs. pure power allocation and pure blocklength allocations. }
\label{pure vs joint new}\vspace{-0.9cm}
\end{figure}

{Finally, we explore the advantage of joint resource allocation design and investigate the respective contributions introduced by the power and the blocklength allocation to system reliability.} As shown in Fig.~\ref{pure vs joint new}, the results of independent power allocation (IPA), independent blocklength allocation (IBA), joint resource allocation (PD) and resource average sharing design (AS) are provided. %, while varying the number of WPT sources. 
{ We assume 
% the added WPT source first toapproaches the center of the circle where the sensors are located and then becomes far away from it, i.e., 
$\rm S_2$ and $\rm S_3$ are relatively closer to sensors while $\rm S_4$, $\rm S_5$ and $\rm S_6$ are relatively further away from sensors.
In Fig.~\ref{pure vs joint new}, we can observe that the joint power and blocklength allocation design significantly outperforms the `AS',`IPA',`IBA' and the performance enhancement enlarges as the source number increases (compared with `AS' and `IBA').
More interestingly, both curve `AS' and `IBA' first decrease and subsequently increase as the source number increases. It can be interpreted by the fact that the new added WPT source $\rm S_4$, $\rm S_5$ and $\rm S_6$ have relatively weaker LoS to sensors compared with $\rm S_2$ and $\rm S_3$. Under total power limitation, power average sharing strategy will diminish the advantage of sources with higher WPT capability and consequently reduce the system FBL reliability. By contrast, `IPA' and `PD' (including power allocation)  can achieve better reliability performance when more sources ($M\geq5$) are deployed for WPT.
Moreover, it can be observed that when $M=1$, the result of `IBA' coincides with that of PD, since the single source takes up the entire power resource without power allocation, i.e., `PD' is equivalent to the `IBA'. Nevertheless, as the number of WPT sources increases, the reliability advantage of `PD' is gradually revealed.
% which indicates that IBA can make relatively more significant contribution to system performance in smaller source number area. 
Besides, by comparing `IPA' and `IBA', we can learn that `IBA' significantly outperforms `IPA' when $M\leq4$, while `IPA' achieves higher reliability when $M\geq5$. The results suggest that in scenarios where power and blocklength are not allowed to be joint optimized simultaneously, pure  power allocation is more proffered for a scenario with more WPT sources, while spending overhead to achieve an efficient  blocklength allocation pays off more when the network has less WPT sources.  }

\vspace{-.25cm}
\section{Conclusion}
%\vspace{-.1cm}
 \label{sec:Conclusion}
In this work, we considered a multi-source WPT enabled multi-sensor communication network, where sensors are activated by the nonlinear EH process and transmit short packets via FBL codes.
For the first time, we characterized the relationship among FBL reliability, multiple source power, WPT blocklength as well as multiple WIT blocklengths for such network.
 With a total source power limitation, we considered a power allocation design aiming at minimizing the overall error probability under fixed frame structure.
{Subsequently, the design  was extended to dynamic frame structure, in which a joint resource allocation design was provided    via jointly optimizing multiple source power  and blocklengths under total power and blocklength constraints.} 
%  Based on characterization, we proposed a power allocation design under fixed frame structure and a joint power and blocklength allocation design under dynamic frame structure with the objective of minimizing the overall error probability.
To solve the challenging non-convex optimization problems, novel variable substitution and SCA algorithm are exploited to reformulate the optimization problems. Based on the iterative algorithm, an efficient sub-optimal solution was thus achieved.
Via numerical simulations, %(did you really use Monte Carlo? why did you use it? This word is not welcome for technical paper, sounds like experimental paper.)}, 
we have demonstrated  significant advantages of our proposed design  and confirmed the importance of the deployment of multiple WPT sources and nonlinear EH model. 
Simulation results also provided guidelines for practical system designs.

%{(Highlighting the contributions and the future extensions)}
At last, we finalize   this work by spotlighting its high extensibility. 
%the high extensibility of our work. (rewrite it? this sentence is the same as example.)
%
% At last, We finalize   this work by reiterating/spotlighting its high extensibility. 
%
%The proposed resource allocation design for minimizing system overall error probability has shown significant benefits of high reliability performance achievement, high design efficiency and accuracy for multiple short packet transmission. 
The methodology of the proposed resource allocation design   minimizing   overall error probability (including the way addressing  the tradeoff among multiple WPT sources)
can be extended to   designs for multi-source WPT enabled  communications, e.g., 
energy-efficiency maximization, end-to-end latency minimization,  location planning for multi WPT sources, and so on.
%the application of this methodology is capable of providing significant energy efficiency enhancement and has potential to be extended to a variety of scenarios, e.g., providing reasonable resource supply strategies for MEC systems, UAV communications and cellular networks and accordingly achieve better system performances, including system reliability, delay, throughput, etc.
{Moreover, the proposed methods for non-convex problem settlement, including variable substitutions, complex relationship decoupling, auxiliary variables introduction, can be applied to facilitate other optimizations of non-convex problems,} especially for scenarios with  coupled  resource  supply and consumption.

\vspace{-.3cm}
\bibliographystyle{IEEEtran}

\bibliographystyle{IEEEtran}

%%%%%%%%%%%%%%%%%%%%%%%%%%%

\end{document}